\documentclass[preprint,aps,onecolumn]{revtex4}%
\usepackage{amssymb}
\usepackage{amsfonts}
\usepackage{amsmath}
\usepackage{graphicx}%
\providecommand{\U}[1]{\protect\rule{.1in}{.1in}}
\newtheorem{theorem}{Theorem}
\newtheorem{acknowledgement}[theorem]{Acknowledgement}

\begin{document}
\preprint{ }
\title[ ]{MESOSCOPIC HYDRO-THERMODYNAMICS of PHONONS}
\author{Aurea R. Vasconcellos$^{(1)}$, A. R. B. de Castro$^{(1,3)}$, C. A. B.
Silva$^{(2)}$}
\author{Roberto Luzzi$^{(1)}$.\thanks{Group Home Page: www.ifi.unicamp.br/ $\thicksim
$aurea \ }}
\affiliation{$(1)$ \textit{Condensed Matter Physics Department}-\textit{Institute of
Physics \textquotedblleft Gleb Wataghin\textquotedblright, State University of
Campinas- Unicamp, 13083-859 Campinas, SP, \ \ Brazil. }}
\affiliation{$(2)$ \textit{Istituto Tecnol\'{o}gico de Aeron\'{a}utica, Departamento de
F\'{\i}sica, 12228-901 S\~{a}o Jos\'{e} dos Campos, SP, Brazil. }}
\affiliation{$(3)$ \textit{Laborat\'{o}rio Nacional de Luz Sincrotron - LNLS, 13083-970
Campinas, SP, Brazil.}}

\begin{abstract}
A generalized Hydrodynamics, referred to as Mesoscopic Hydro-Thermodynamics,
of phonons in semiconductors is presented. It involves the descriptions of the
motion of the quasi-particle density and of the energy density. The
hydrodynamic equations, which couple both types of movement via thermo-elastic
processes, are derived starting with a generalized Peierls-Boltzmann kinetic
equation obtained in the framework of a Non-Equilibrium Statistical Ensemble
Formalism, providing such Mesoscopic Hydro-Thermodynamics.

The case of \ a contraction in first order is worked out in detail. The
associated Maxwell times are derived and discussed. The densities of
quasi-particles and of energy are found to satisfy coupled
Maxwell-Cattaneo-like (hyperbolic) equations. The analysis of thermo-elastic
effects is done and applied to investigate thermal distortion in silicon
mirrors under incidence of high intensity X-ray pulses in FEL facilities. The
derivation of a generalized Guyer-Krumhansl equation governing the flux of
heat and the associated thermal conductivity cofficient is also presented.

\end{abstract}
\maketitle

\section{Introduction}

It has been noticed \cite{editors} that the ceaseless innovation in
semiconductors design creates a demand for better understanding of the
physical processes involved in the functioning of modern electronic devices,
which operate under \ far-from-equilibrium conditions. The main interest is
centered on the behavior of \textquotedblleft hot\textquotedblright\ carriers
(electrons and holes) but the case of \textquotedblleft hot\textquotedblright%
\ phonons is also of interest, particularly in questions such as refrigeration
of microprocessors and heat transport in small systems with constrained
geometries [\cite{zhang}-\cite{chowdhury}].

These questions belong \ to the area of non-equilibrium phonon-dynamics
\cite{klein}, more precisely to the subject of phonon hydrodynamics associated
to non-equilibrium (irreversible) thermodynamics [\cite{jou}, \cite{luzzi}].

It may be noticed the fact that kinetic and hydrodynamics of fluids are
intimately coupled. A first kinetic-hydrodynamic approach can be considered to
be the so-called classical (or Onsagerian) hydrodynamics, which gives
microscopic (mechano-statistical) foundations to, for example, the classical
Fourier's and Fick's diffusion laws. But it works under quite restrictive
conditions, (see for example \cite{kreuzer}) and, hence, more advanced
approaches are required to lift these restrictions, mainly, because, as
noticed, the requirements for the analysis of situations involved in the
nowadays advanced technologies. Several improved approaches were introduced
as, for example, the Burnett approximation of hydrodynamics \cite{mackowski},
the one associated to Extended Irreversible Thermodynamics \cite{jou}, and,
recently, the so called Mesoscopic Hydro-Thermodynamics (MHT for short)
dealing with heat transport \cite{dende} covering, in principle, all kinds of
motion, that is, involving intermediate to short wavelengths and ultrafast
motion. A complete MHT for classical fluids is reported in Ref. \cite{silva}.

In this communication we describe the construction of a MHT of phonons,
presenting an in depth study of phonon hydro-thermodynamics in the framework
of a nonlinear quantum kinetic theory based on a non-equilibrium statistical
ensemble formalism [\cite{akhiezer}-\cite{vannucchi}], referred to as NESEF
for short.

In \textbf{section II }derivation of a phonon higher-order hydrodynamics is
presented. The hydrodynamic equations consist of the coupled sets of
time-evolution equations for (a) the density \ of quasi-particles, together
with those for its fluxes of all orders, and (b) the density of energy and its
fluxes (heat transport) of all orders. The two families of equations are
coupled together through thermo-elastic effects. The practical application of
such encumbered sets of equations requires to introduce a contracted
description, that is, to \ retain only a finite number of fluxes and closing
the set of equations by means of auxiliary complementary equations.

In \textbf{section III} we fully derive a description said of order one, where
we retain only the two densities and the first flux of each density.

In \textbf{section IV} we present an analysis of some characteristics of these
equations, namely, (a) the Maxwell times \cite{maxwell} associated to the
densities and fluxes, and (b) the construction of a hyperbolic
Maxwell-Cattaneo-like equations for both densities and of a
Guyer-Krumhansl-like equation for the heat flux.

In \textbf{section V} we consider the simpler case when thermo-elastic effects
are neglected and the equations for the two densities decouple.

In \textbf{section VI}, on the contrary, thermo-elastic effects are analyzed
and used to investigate transient thermal distortion of an optical substrate
illuminated by a single intense ultra-short X-ray pulse as currently available
in FEL facilities.

\textbf{Section VII} contains our concluding remarks. Details of the
calculation are given in several \textbf{Appendices}.

\section{Phonon Mesoscopic Hydro-Thermodynamics}

We consider a system of acoustic phonons in a semiconductor, in anharmonic
interaction among them, in interaction with what we call a thermal bath of
other degrees of freedom of the system, and in the presence of an external
source capable of driving them out of thermal equilibrium. Moreover, the
system is in contact with an external thermostat at temperature $T_{0}$. Since
the divergence of a transverse field vanishes, and since in the absence of
vortices the rotational also vanishes, in what regards the
\textit{hydrodynamic motion} we consider the longitudinal acoustic (LA)
phonons only.

\qquad The system Hamiltonian quantum mechanical operator is%
\begin{equation}
\widehat{H}=\widehat{H}_{OS}+\widehat{H}_{SB}+\widehat{H}_{SP}+\widehat{H}%
_{OB} \label{eq1}%
\end{equation}
where $\widehat{H}_{OS}$ is the Hamiltonian operator of the free LA system,
namely,\qquad%
\begin{equation}
\widehat{H}_{OS}=%
{\textstyle\sum\nolimits_{\mathbf{q}}}
\hslash\omega_{\mathbf{q}}\left(  \widehat{a}_{\mathbf{q}}^{+}\widehat{a}%
_{\mathbf{q}}+1/2\right)  \label{eq2}%
\end{equation}
with $\omega_{\mathbf{q}}$ being the frequency dispersion relation, the
wavevector $\mathbf{q}$ runs over the 1st Brillouin zone and $\widehat{a}%
_{\mathbf{q}}$, $\widehat{a}_{\mathbf{q}}^{+}$ are annihilation and creation
operators for LA phonons in mode $\mathbf{q}$.

The $\widehat{H}_{SB}$ term accounts for the interaction of the LA phonons
with what we have called the thermal bath, namely, the anharmonic interaction
with the TA phonons, the effect of impurities elastic deformation and
interaction with the electrons, or holes, they being, bonded or itinerant, the
effect of imperfections (mainly dislocations), and we write for it
\begin{equation}
\widehat{H}_{SB}=%
{\textstyle\sum\nolimits_{\mathbf{q}}}
\left(  \Lambda_{\mathbf{q}}\widehat{R}_{\mathbf{q}}\widehat{a}_{\mathbf{q}%
}^{+}+\Lambda_{\mathbf{q}}^{\ast}\widehat{R}_{\mathbf{q}}^{+}\widehat{a}%
_{\mathbf{q}}\right)  \label{eq3}%
\end{equation}
where $\widehat{R}_{\mathbf{q}}$ and $\widehat{R}_{\mathbf{q}}^{+}$ indicate
the corresponding operators associated with transition processes involving the
interactions mentioned above; $\Lambda_{\mathbf{q}}$ is the strength coupling
in each case. For illustration let us consider the anharmonic interaction
operator between the LA and TA phonons, namely$\ $%
\begin{align}
\widehat{H}_{SB}  &  =%
{\textstyle\sum\nolimits_{\mathbf{qk}}}
\left\{  M_{\mathbf{qk}}\widehat{a}_{\mathbf{q}}\widehat{b}_{\mathbf{q-k}}%
^{+}\widehat{b}_{\mathbf{-k}}^{+}+M_{\mathbf{qk}}^{^{\prime}}\widehat{a}%
_{\mathbf{q}}\widehat{a}_{\mathbf{q-k}}^{+}\widehat{b}_{\mathbf{-k}}%
^{+}+M_{\mathbf{qk}}^{\prime\prime}\widehat{a}_{\mathbf{q}}\widehat{a}%
_{\mathbf{q-k}}\widehat{b}_{\mathbf{k}}^{+}\right\} \nonumber\\
&  +\text{ }Hermitian\text{ }conjugate, \label{eq4}%
\end{align}
where the first contribution on the RHS describes the processes $LA(\mathbf{q}%
)\rightarrow TA(\mathbf{q}^{\prime})+TA(\mathbf{q}^{\prime\prime})$, the
second the processes $LA(\mathbf{q})\rightarrow LA(\mathbf{q}^{\prime
})+TA(\mathbf{q}^{\prime\prime})$, the third $LA(\mathbf{q})+LA(\mathbf{q}%
^{\prime})\rightarrow TA(\mathbf{q}^{\prime\prime})$ and the Hermitian
conjugate describe the inverse processes. $M_{\mathbf{qk}}$ are the
appropriate matrix elements. The operators $\widehat{a}_{\mathbf{q}%
}\widehat{b}_{\mathbf{q-k}}^{+}\widehat{b}_{\mathbf{-k}}^{+}$ lead to the
appearance, in the kinetic equation of evolution, of a linear term
characterized by a relaxation time $\tau_{\mathbf{q}}^{\mathbf{an}}$, while
the operators $\widehat{a}_{\mathbf{q}}\widehat{a}_{\mathbf{q-k}}%
^{+}\widehat{b}_{\mathbf{-k}}^{+}$ and $\widehat{a}_{\mathbf{q}}%
\widehat{a}_{\mathbf{q-k}}\widehat{b}_{\mathbf{k}}^{+}$ produce non-linear
terms sometimes referred to as Lifshits \cite{livshits} and Fr\"{o}hlich
\cite{frolich} contributions, respectively. It is worth mentioning that these
non-linear terms produce, when the system is driven (by the external pumping
source) sufficiently far from equilibrium, a so called \textquotedblleft
complex behavior\textquotedblright\ [\cite{haken}, \cite{nicolis}]. In the
case of phonons, it consists in the emergence of a type of Bose-Einstein
condensation, propagation of long-lived solitons and a kind of
\textquotedblleft laser\textquotedblright\ action [\cite{frolich1}%
-\cite{rodrigues1}]. In what follows we shall take the case of not too intense
excitation and disregard non-linear contributions, restricting the analysis to
the linear (Onsagerian) regime, when the emergence of \textquotedblleft
complex behavior\textquotedblright\ is restrained according to Prigogine's
theorem of minimum entropy production in local equilibrium \cite{glansdorff}.

\qquad Moreover, $\widehat{H}_{OB}$ is the Hamiltonian of the subsystems
involved in the thermal bath, and $\widehat{H}_{SP}$ is the interaction of the
LA-phonons with the external pumping source, to be specified in each practical application.

\qquad Next, to apply NESEF it is required first of all to specify the basic
dynamic variables used to characterize the non-equilibrium ensemble
[\cite{zubarev}-\cite{vannucchi}]. A priori, when the system is driven away
from equilibrium, it is necessary to include all observables of the system,
which leads to the introduction of many-particle dynamic operators
[\cite{fano}, \cite{bogoliubov}]. Here we use the single-particle dynamic
operator%
\begin{equation}
\widehat{\nu}_{\mathbf{qQ}}=\widehat{a}_{\mathbf{q+Q/2}}^{+}\widehat{a}%
_{\mathbf{q-Q/2}}\label{eq5}%
\end{equation}
in the second quantization representation in reciprocal space, with \textbf{q}
and \textbf{Q} running over the Brillouin zone.

\qquad The two-phonon and higher-order dynamic operators can be ignored
because of Bogoliubov's principle of correlation weakening [\cite{bogoliubov1}%
, \cite{uhlenbeck}]. Moreover, since phonons are bosons, it would be necessary
also to include the creation and annihilation operators $\widehat{a}%
_{\mathbf{q}}$ and $\widehat{a}_{\mathbf{q}}^{+}$ because their eigenstates
are the coherent states \cite{klauder}, and also the pair operators
$\widehat{a}_{\mathbf{k}}\widehat{a}_{\mathbf{k}^{^{,}}}$ ($\widehat{a}%
_{\mathbf{k}}^{+}\widehat{a}_{\mathbf{k}^{,}}^{+}$) because the number of
quasi-particles is not fixed \cite{hugenholtz}. However, we disregard them
because they are of no practical relevance for the problem at hands. In
\textbf{Appendix A} we describe the corresponding non-equilibrium statistical
operator. The energy of the thermal bath $\widehat{H}_{OB}$ is also a basic
microdynamical variable and then we have the basic set%
\begin{equation}
\left\{  \widehat{\nu}_{\mathbf{qQ}},\text{ }\widehat{H}_{OB}\right\}
,\label{eq6}%
\end{equation}
where, $\mathbf{Q}=0$ corresponds to the occupation number operator
$\widehat{\nu}_{\mathbf{q}}=\widehat{a}_{\mathbf{q}}^{+}\widehat{a}%
_{\mathbf{q}}$ describing a homogeneous phonon population and those with
$\mathbf{Q}\neq0$, namely $\widehat{\nu}_{\mathbf{qQ}}=\widehat{a}%
_{\mathbf{q+Q/2}}^{+}\widehat{a}_{\mathbf{q-Q/2}}$, account for changes in
space of the non-equilibrium phonon distribution function.

\qquad The average, over the non-equilibrium ensemble, of the microdynamical
variables in the set of Eq.(\ref{eq6}) provides the variables which
characterize the non-equilibrium macroscopic state of the system. Let us call
them%
\begin{equation}
\{\nu_{\mathbf{q}}(t)=<\widehat{\nu}_{\mathbf{q}}|t>;\text{ \ }\nu
_{\mathbf{qQ}}(t)=<\widehat{\nu}_{\mathbf{qQ}}|t>;\text{ \ }E_{B}%
=<\widehat{H}_{OB}>\} \label{eq7}%
\end{equation}
where $\mathbf{Q}\neq0$ and $<\widehat{\nu}_{\mathbf{q}}|t>=Tr\{\widehat{\nu
}_{\mathbf{q}}\varrho_{\varepsilon}\left(  t\right)  \}$, etc..., that is, the
average of the microdynamical operators of the set (\ref{eq6}) over the
non-equilibrium ensemble according to the formalism in \textbf{Appendix A},
where we have introduced the non-equilibrium thermodynamic state variables
conjugated to those above, namely [cf. Eq.(\ref{eqA3})]%
\begin{equation}
\{F_{\mathbf{q}}(t),\text{ }F_{\mathbf{qQ}}(t),\text{ }\beta_{0}\};
\label{eq8}%
\end{equation}
where $1/\beta_{0}=k_{B}T_{0}$.

\qquad Going over to direct space we introduce the space and crystal momentum
dependent distribution function%
\begin{equation}
\nu_{\mathbf{q}}(\mathbf{r},t)=\frac{1}{V}%
{\textstyle\sum\nolimits_{\mathbf{Q}}}
\nu_{\mathbf{qQ}}(t)exp[i\mathbf{Q}\cdot\mathbf{r}], \label{eq9}%
\end{equation}
$V$ is the sample volume which shall be taken as one in what follows, where it
may be noticed that $\nu_{\mathbf{q}}(t)$, $\mathbf{Q}=0$, refers to the
global properties and those with $\mathbf{Q}\neq0$ reflect the changes in
direct space. Hence, the hydrodynamic variables that describe the hydrodynamic
movement consist of two families, namely%
\begin{equation}
\{n(\mathbf{r},t),\text{ }\mathbf{I}_{n}(\mathbf{r},t\mathbf{),...,}\text{
}I_{n}^{[\ell]}(\mathbf{r},t\mathbf{),...\}} \label{eq10}%
\end{equation}
with $\ell=2,3,\ldots,$\ describing the \textit{movement of the
quasi-particles, which we call the n-family}, and where%
\begin{equation}
n(\mathbf{r},t)=%
{\textstyle\sum\nolimits_{\mathbf{q}}}
\nu_{\mathbf{q}}(\mathbf{r},t) \label{eq11}%
\end{equation}
is the density of the quasi-particles,%
\begin{equation}
\mathbf{I}_{n}(\mathbf{r},t\mathbf{)=}%
{\textstyle\sum\nolimits_{\mathbf{q}}}
\nabla_{\mathbf{q}}\omega_{\mathbf{q}}\text{ }\nu_{\mathbf{q}}(\mathbf{r},t)
\label{eq12}%
\end{equation}
is the first (vectorial) flux of the density (current),%
\begin{equation}
I_{n}^{[\ell]}(\mathbf{r},t\mathbf{)=}%
{\textstyle\sum\nolimits_{\mathbf{q}}}
\nabla_{\mathbf{q}}^{\left[  \ell\right]  }\omega_{\mathbf{q}}\text{ }%
\nu_{\mathbf{q}}(\mathbf{r},t) \label{eq13}%
\end{equation}
is the rank-$\ell$, with $\ell>1$, ($\ell=2,3,4\ldots$) tensorial flux of the
density, where%
\begin{equation}
\nabla_{\mathbf{q}}^{\left[  \ell\right]  }\omega_{\mathbf{q}}=\left[
\nabla_{\mathbf{q}}\omega_{\mathbf{q}}\nabla_{\mathbf{q}}\omega_{\mathbf{q}%
}...\ell\text{ }times...\text{ }\nabla_{\mathbf{q}}\omega_{\mathbf{q}}\right]
\label{eq15}%
\end{equation}
meaning tensorial product of $\ell$-times the vector $\nabla_{\mathbf{q}%
}\omega_{\mathbf{q}}$, rendering a tensor of $\ell$-rank.

\qquad On the other hand, we have%
\begin{equation}
\{h(\mathbf{r},t),\text{ }\mathbf{I}_{h}(\mathbf{r},t\mathbf{),...,}\text{
}I_{h}^{[\ell]}(\mathbf{r},t\mathbf{),...\}} \label{eq16}%
\end{equation}
again with $\ell=2,3,\ldots$, describing \textit{the movement of energy of
quasi-particles, which we call the h-family}, and where%
\begin{equation}
h(\mathbf{r},t)=%
{\textstyle\sum\nolimits_{\mathbf{q}}}
\hslash\omega_{\mathbf{q}}\nu_{\mathbf{q}}(\mathbf{r},t), \label{eq17}%
\end{equation}%
\begin{equation}
\mathbf{I}_{h}(\mathbf{r},t\mathbf{)=}%
{\textstyle\sum\nolimits_{\mathbf{q}}}
\hslash\omega_{\mathbf{q}}\nabla_{\mathbf{q}}\omega_{\mathbf{q}}\text{ }%
\nu_{\mathbf{q}}(\mathbf{r},t), \label{eq18}%
\end{equation}%
\begin{equation}
I_{h}^{[\ell]}(\mathbf{r},t\mathbf{)=}%
{\textstyle\sum\nolimits_{\mathbf{q}}}
\hslash\omega_{\mathbf{q}}\nabla_{\mathbf{q}}^{\left[  \ell\right]  }%
\omega_{\mathbf{q}}\text{ }\nu_{\mathbf{q}}(\mathbf{r},t), \label{eq19}%
\end{equation}
which are, respectively, the density of energy, its first flux, and the higher
order fluxes.

\qquad By deriving on time $t$, the two sides of the general equations
(\ref{eq11}) to (\ref{eq13}) and (\ref{eq17}) to (\ref{eq19}), there follow,
written in compact form, the hydrodynamic equations of motion%
\begin{equation}
\frac{\partial}{\partial t}I_{p}^{[\ell]}(\mathbf{r},t)=%
{\textstyle\sum\nolimits_{\mathbf{q}}}
K_{p}^{[\ell]}(\mathbf{q})\frac{\partial}{\partial t}\nu_{\mathbf{q}%
}(\mathbf{r},t) \label{eq20}%
\end{equation}
where $p$ is $n$ or $h$ (for the corresponding families), and%
\begin{equation}
K_{n}^{[\ell]}(\mathbf{q})=\nabla_{\mathbf{q}}^{[\ell]}\omega_{\mathbf{q}},
\label{eq21}%
\end{equation}
and,%
\begin{equation}
K_{h}^{[\ell]}(\mathbf{q})=\hslash\omega_{\mathbf{q}}\nabla_{\mathbf{q}%
}^{\left[  \ell\right]  }\omega_{\mathbf{q}}. \label{eq22}%
\end{equation}

\qquad Evidently, all the hydrodynamic equations of motion are dependent on
the equation of motion for the single quantity $\nu_{\mathbf{q}}%
(\mathbf{r},t)$. Returning, for practical convenience, to reciprocal
$\mathbf{Q}$ space, $\nu_{\mathbf{qQ}}(t)$ satisfies the evolution equation%
\begin{equation}
\frac{\partial}{\partial t}\nu_{\mathbf{qQ}}(t)=Tr\{(i\hslash)^{-1}%
[\widehat{\nu}_{\mathbf{qQ}},\widehat{H}]\varrho_{\varepsilon}\left(
t\right)  \} \label{eq23}%
\end{equation}
which is the average over the non-equilibrium ensemble (described by the
statistical operator $\varrho_{\varepsilon}\left(  t\right)  $), of the
quantum-mechanical Heisenberg equation of motion for the microdynamical
variable (operator) $\widehat{\nu}_{\mathbf{qQ}}$ given in Eq.(\ref{eq5}).

\qquad Direct calculation of the RHS in Eq.(\ref{eq23}) is extremely difficult
and then it is necessary to resort to the introduction of a more practical
\textit{non-linear quantum kinetic theory} [\cite{akhiezer}-\cite{vannucchi}]
briefly described in \textbf{Appendix B}, which is applied using an
approximation consisting in retaining only the collision integral which is
second order in the interaction strength. The resulting evolution equation,
when rewritten in direct \textbf{r}-space becomes a generalization of the
Peierls-Boltzmann kinetic equation. As shown in\textbf{\ Appendix B}, once the
limit of large wavelengths is taken, it acquires a form resembling the
standard one, to be used consistently in what follows, namely%
\[
\frac{\partial}{\partial t}\nu_{\mathbf{q}}(\mathbf{r},t)=\nabla_{\mathbf{q}%
}\omega_{\mathbf{q}}\cdot\nabla_{\mathbf{r}}\nu_{\mathbf{q}}(\mathbf{r}%
,t)+\nabla_{\mathbf{q}}\Pi_{\mathbf{q}}\cdot\nabla_{\mathbf{r}}\nu
_{\mathbf{q}}(\mathbf{r},t)
\]%
\begin{equation}
-\Gamma_{\mathbf{q}}\left[  \nu_{\mathbf{q}}(\mathbf{r},t)-\nu_{\mathbf{q}%
}^{\mathbf{0}}\right]  +\mathcal{I}_{\mathbf{q}}^{\mathbf{ext.}}\left(
\mathbf{r},t\right)  \label{eq24}%
\end{equation}
with%
\begin{equation}
\nu_{\mathbf{q}}^{\mathbf{0}}=1/\left[  exp\left(  \hslash\omega_{\mathbf{q}%
}/k_{B}T_{0}\right)  -1\right]  \label{eq25}%
\end{equation}
where $\nu_{\mathbf{q}}^{\mathbf{0}}$ is the equilibrium LA phonon
distribution at temperature $T_{0}$, $\mathcal{I}_{\mathbf{q}}^{\mathbf{ext.}%
}$ describes the rate of \ change due to external sources/sinks acting on the
system, $\Gamma_{\mathbf{q}}$ is a reciprocal lifetime given in Eq.(\ref{eq26}%
), $\Pi_{\mathbf{q}}$ is the self-energy correction of the LA phonon frequency
giving $\overline{\omega}_{\mathbf{q}}$ of Eq.(\ref{eq28}), which, if in
Eq.(\ref{eq3}) we consider only the linear anharmonic interaction, that is,
the first term on the right of Eq.(\ref{eq4}) and its Hermitian conjugate, are
given by
\begin{equation}
\Gamma_{\mathbf{q}}=\pi/\hslash^{2}%
{\textstyle\sum\nolimits_{\mathbf{k}}}
\left\vert M_{\mathbf{kq}}\right\vert ^{2}\left(  1+\nu_{\mathbf{k}%
}^{\mathbf{TA}}+\nu_{\mathbf{k+q}}^{\mathbf{TA}}\right)  \delta\left(
\Omega_{\mathbf{k+q}}+\Omega_{\mathbf{k}}-\omega_{\mathbf{q}}\right)
,\label{eq26}%
\end{equation}%
\begin{equation}
\nu_{\mathbf{k}}^{\mathbf{TA}}=1/\left[  exp\left(  \hslash\Omega_{\mathbf{k}%
}/k_{B}T_{0}\right)  -1\right]  ,\label{eq27}%
\end{equation}%
\begin{equation}
\overline{\omega}_{\mathbf{q}}=\omega_{\mathbf{q}}+\Pi_{\mathbf{q}%
},\label{eq28}%
\end{equation}%
\begin{equation}
\Pi_{\mathbf{q}}=(\pi/\hslash^{2})P.V.%
{\textstyle\sum\nolimits_{\mathbf{k}}}
\left\vert M_{\mathbf{kq}}\right\vert ^{2}\left(  1+\nu_{\mathbf{k}%
}^{\mathbf{TA}}+\nu_{\mathbf{k+q}}^{\mathbf{TA}}\right)  /\left(
\Omega_{\mathbf{k+q}}+\Omega_{\mathbf{k}}-\omega_{\mathbf{q}}\right)
,\label{eq29}%
\end{equation}
where $\Omega_{\mathbf{k}}$ is the $\mathbf{k}$-dependent frequency of TA
phonons whose distribution is $\nu_{\mathbf{k}}^{\mathbf{TA}}$, and $P.V.$
stands for principal value.

\qquad Inserting Eq.(\ref{eq24}) in Eq.(\ref{eq20}), the hydrodynamic
evolution equations for the \textit{n} and \textit{h} families become%
\[
\frac{\partial}{\partial t}I_{p}^{[\ell]}(\mathbf{r},t)=%
{\textstyle\sum\nolimits_{\mathbf{q}}}
K_{p}^{[\ell]}(\mathbf{q})\left\{  \nabla_{\mathbf{q}}\omega_{\mathbf{q}}%
\cdot\nabla_{\mathbf{r}}\nu_{\mathbf{q}}(\mathbf{r},t)+\nabla_{\mathbf{q}}%
\Pi_{\mathbf{q}}\cdot\nabla_{\mathbf{r}}\nu_{\mathbf{q}}(\mathbf{r},t)\right.
\]%
\begin{equation}
\left.  -\Gamma_{\mathbf{q}}\left[  \nu_{\mathbf{q}}(\mathbf{r},t)-\nu
_{\mathbf{q}}^{\mathbf{0}}\right]  +\mathcal{I}_{\mathbf{q}}^{\mathbf{ext.}%
}\left(  \mathbf{r},t\right)  \right\}  . \label{eq30}%
\end{equation}

\qquad Next, a closure for the set of Eq.(\ref{eq30}) must be introduced. This
means, that we must express the $\nu_{\mathbf{q}}(\mathbf{r},t)$ appearing on
the RHS in terms of the hydrodynamical variables belonging to the set. First
thing to notice is that we are dealing with an enormous set of coupled
integro-differential equations linking densities and fluxes of all orders. In
reference \cite{madureira} (equation 7) an equivalent representation, in
direct $\mathbf{r}$-space, of these equations is given as%
\begin{equation}
\frac{\partial}{\partial t}I_{p}^{[\ell]}(\mathbf{r},t)+\nabla_{\mathbf{r}%
}\cdot I_{p}^{[\ell+1]}(\mathbf{r},t)=%
{\textstyle\sum\nolimits_{p\acute{}\ell\acute{}}}
{\textstyle\int}
d^{3}\mathbf{r}%
\acute{}%
\mathcal{L}_{pp%
\acute{}%
}^{\left[  \ell+\ell%
\acute{}%
\right]  }\left(  \mathbf{r}-\mathbf{r}%
\acute{}%
,t\right)  \odot I_{p%
\acute{}%
}^{[\ell%
\acute{}%
]}(\mathbf{r}%
\acute{}%
,t)+\mathcal{I}_{p}^{\left[  \ell\right]  \mathbf{ext.}}\left(  \mathbf{r}%
,t\right)  , \label{eq31}%
\end{equation}
where $\mathcal{L}_{pp%
\acute{}%
}^{\left[  \ell+\ell%
\acute{}%
\right]  }\left(  \mathbf{r}-\mathbf{r}%
\acute{}%
,t\right)  \odot I_{p%
\acute{}%
}^{[\ell%
\acute{}%
]}(\mathbf{r}%
\acute{}%
,t)$ stands for the contraction in all $\ell%
\acute{}%
$ tensor indices.

\qquad Equation (\ref{eq31}) is a continuity equation, having on the RHS a
collision integral which accounts for sinks (system relaxation effects) and
sources (pumping, driving the system out of equilibrium). It can be considered
an extended Mori-Heisenberg-Langevin equation \cite{mori}. It represents a
quite cumbersome set of coupled equations, of unmanageable proportion. To
proceed further it is then necessary to introduce a \textit{contraction of
description} \cite{ramos}, that is to say, a reduction in the number of basic
quantities, retaining only a few fluxes. Hence, we must look, in each case, on
how to find the best description using the smallest possible number of
variables. In other words to introduce an appropriate \textit{contraction of
description:} \textit{This contraction implies in retaining the information
considered as relevant for the problem in hands, and to disregard nonrelevant
information} \cite{balian}.

Elsewhere \cite{ramos} we have considered the question of the contraction of
description (reduction of dimensions of the nonequilibrium thermodynamic space
of states), where a criterion for justifying the different levels of
truncation is derived: It depends on the range of wavelengths and frequencies
which are relevant for the characterization, in terms of normal modes, of the
hydro-thermodynamic motion in the nonequilibrium open system.

In other words, since MHT implies in describing the motion when governed by
smaller and smaller wavelengths, or larger and larger wavenumbers, accompanied
by higher and higher frequencies, in a qualitative manner we can say that, as
a general \ \textquotedblleft thumb rule,\textquotedblright\ the criterion
indicates that \textit{a more and more restricted contraction can be used when
larger and larger are the prevalent wavelengths in the motion (changes
smoother and smoother in space and time). }Therefore, in simpler words, when
the motion becomes more and more smooth in space and time, the more reduced
can be the dimension of the basic macrovariables space to be used for the
description of the nonequilibrium thermodynamic state of the system.

It can be conjectured \ a general criterion for performing contractions,
namely, a contraction of order \textit{r} (meaning keeping the densities and
their fluxes \ up to order \textit{r}) can be introduced, once we can show
that in the spectrum of wavelengths, which characterize the motion,
predominate those larger than a \textquotedblleft frontier\textquotedblright%
\ one, $\lambda_{\left(  r,r+1\right)  }^{2}=v^{2}\theta_{r}\theta_{r+1}$,
where $v$ is of the order of the thermal velocity and $\theta_{r}$ and
$\theta_{r+1}$ the corresponding Maxwell times \cite{ramos}.

\qquad In the next section we consider the situation when it is possible to
use a contracted description including only the densities and their first fluxes.

\section{Analysis of the Phonon MHT of Order 1}

Let us consider a contracted description including the densities of
quasi-particles $n(\mathbf{r},t)$ and energy $h(\mathbf{r},t)$ plus their
first fluxes $\mathbf{I}_{n}(\mathbf{r},t\mathbf{)}$, $\mathbf{I}%
_{h}(\mathbf{r},t\mathbf{)}$ only. For practical convenience it is better to
work in reciprocal $\mathbf{Q}$ space. These quantities are then given by%
\begin{equation}
n(\mathbf{Q},t)=%
{\textstyle\sum\nolimits_{\mathbf{q}}}
\nu_{\mathbf{qQ}}(t), \label{eq32}%
\end{equation}%
\begin{equation}
h(\mathbf{Q},t)=%
{\textstyle\sum\nolimits_{\mathbf{q}}}
\hslash\omega_{\mathbf{q}}\text{ }\nu_{\mathbf{qQ}}(t), \label{eq33}%
\end{equation}%
\begin{equation}
\mathbf{I}_{n}(\mathbf{Q},t\mathbf{)=}%
{\textstyle\sum\nolimits_{\mathbf{q}}}
\nabla_{\mathbf{q}}\omega_{\mathbf{q}}\text{ }\nu_{\mathbf{qQ}}(t),
\label{eq34}%
\end{equation}%
\begin{equation}
\mathbf{I}_{h}(\mathbf{Q},t\mathbf{)=}%
{\textstyle\sum\nolimits_{\mathbf{q}}}
\hslash\omega_{\mathbf{q}}\nabla_{\mathbf{q}}\omega_{\mathbf{q}}\text{ }%
\nu_{\mathbf{qQ}}(t). \label{eq35}%
\end{equation}

\qquad It is recalled that%
\begin{equation}
\nu_{\mathbf{qQ}}\left(  t\right)  =Tr\left\{  \widehat{a}_{\mathbf{q+Q/2}%
}^{+}\widehat{a}_{\mathbf{q-Q/2}}\text{ }\overline{\varrho}\left(  t,0\right)
\right\}  , \label{eq36}%
\end{equation}
where (see \textbf{Appendix A})%

\begin{equation}
\overline{\varrho}\left(  t,0\right)  =exp\{-\phi(t)-%
{\textstyle\sum\nolimits_{\mathbf{q}}}
\left[  F_{\mathbf{q}}\left(  t\right)  \widehat{\nu}_{\mathbf{q}}+%
{\textstyle\sum\nolimits_{\mathbf{Q\neq0}}}
F_{\mathbf{qQ}}\left(  t\right)  \widehat{\nu}_{\mathbf{qQ}}\right]  \}
\label{eq37}%
\end{equation}
with, in this contracted description,%
\begin{equation}
F_{\mathbf{qQ}}\left(  t\right)  =\varphi_{n}\left(  \mathbf{Q},t\right)
+\varphi_{h}\left(  \mathbf{Q},t\right)  \hslash\omega_{\mathbf{q}}%
+\mathbf{F}_{n}\left(  \mathbf{Q},t\right)  \mathbf{\cdot}\nabla_{\mathbf{q}%
}\omega_{\mathbf{q}}+\mathbf{F}_{h}\left(  \mathbf{Q},t\right)  \mathbf{\cdot
}\hslash\omega_{\mathbf{q}}\nabla_{\mathbf{q}}\omega_{\mathbf{q}}.
\label{eq38}%
\end{equation}

\qquad Using the generalized Peierls-Boltzmann Eq.(\ref{eqB6}), the resulting
equations of motion are ($\mathbf{Q\neq0}$)
\[
\partial n(\mathbf{Q},t)/\partial t=i\mathbf{Q\cdot I}_{n}\mathbf{(Q,}%
t\mathbf{)-}\left(  i/2\right)
{\textstyle\sum\nolimits_{\mathbf{q}}}
\left(  \Pi_{\mathbf{q+Q/2}}-\Pi_{\mathbf{q-Q/2}}\right)  \nu_{\mathbf{qQ}%
}(t)\text{ }%
\]%
\begin{equation}
\mathbf{-}\left(  1/2\right)
{\textstyle\sum\nolimits_{\mathbf{q}}}
\left(  \Gamma_{\mathbf{q+Q/2}}+\Gamma_{\mathbf{q-Q/2}}\right)  \nu
_{\mathbf{qQ}}(t)+\mathcal{I}_{n}^{\left[  0\right]  \mathbf{ext.}}\left(
\mathbf{Q},t\right)  , \label{eq39}%
\end{equation}%
\[
\partial h(\mathbf{Q},t)/\partial t=i\mathbf{Q\cdot I}_{h}\mathbf{(Q,}%
t\mathbf{)-}\left(  i/2\right)
{\textstyle\sum\nolimits_{\mathbf{q}}}
\left(  \Pi_{\mathbf{q+Q/2}}-\Pi_{\mathbf{q-Q/2}}\right)  \nu_{\mathbf{qQ}%
}(t)\hslash\omega_{\mathbf{q}}%
\]%
\begin{equation}
\mathbf{-}\left(  1/2\right)
{\textstyle\sum\nolimits_{\mathbf{q}}}
\left(  \Gamma_{\mathbf{q+Q/2}}+\Gamma_{\mathbf{q-Q/2}}\right)  \nu
_{\mathbf{qQ}}(t)\hslash\omega_{\mathbf{q}}+\mathcal{I}_{h}^{\left[  0\right]
\mathbf{ext.}}\left(  \mathbf{Q},t\right)  , \label{eq40}%
\end{equation}%
\[
\partial\mathbf{I}_{n}(\mathbf{Q},t\mathbf{)/\partial}t=i\mathbf{Q\cdot}%
I_{n}^{\left[  2\right]  }(\mathbf{Q},t\mathbf{)-}\left(  i/2\right)
{\textstyle\sum\nolimits_{\mathbf{q}}}
\left(  \Pi_{\mathbf{q+Q/2}}-\Pi_{\mathbf{q-Q/2}}\right)  \nu_{\mathbf{qQ}%
}(t)\nabla_{\mathbf{q}}\omega_{\mathbf{q}}%
\]%
\begin{equation}
\mathbf{-}\left(  1/2\right)
{\textstyle\sum\nolimits_{\mathbf{q}}}
\left(  \Gamma_{\mathbf{q+Q/2}}+\Gamma_{\mathbf{q-Q/2}}\right)  \nu
_{\mathbf{qQ}}(t)\nabla_{\mathbf{q}}\omega_{\mathbf{q}}+\mathcal{I}%
_{n}^{\left[  1\right]  \mathbf{ext.}}\left(  \mathbf{Q},t\right)  ,
\label{eq41}%
\end{equation}%
\[
\partial\mathbf{I}_{h}(\mathbf{Q},t\mathbf{)/\partial}t=i\mathbf{Q\cdot}%
I_{h}^{\left[  2\right]  }(\mathbf{Q},t\mathbf{)-}\left(  i/2\right)
{\textstyle\sum\nolimits_{\mathbf{q}}}
\left(  \Pi_{\mathbf{q+Q/2}}-\Pi_{\mathbf{q-Q/2}}\right)  \nu_{\mathbf{qQ}%
}(t)\hslash\omega_{\mathbf{q}}\nabla_{\mathbf{q}}\omega_{\mathbf{q}}%
\]%
\begin{equation}
\mathbf{-}\left(  1/2\right)
{\textstyle\sum\nolimits_{\mathbf{q}}}
\left(  \Gamma_{\mathbf{q+Q/2}}+\Gamma_{\mathbf{q-Q/2}}\right)  \nu
_{\mathbf{qQ}}(t)\hslash\omega_{\mathbf{q}}\nabla_{\mathbf{q}}\omega
_{\mathbf{q}}+\mathcal{I}_{h}^{\left[  1\right]  \mathbf{ext.}}\left(
\mathbf{Q},t\right)  . \label{eq42}%
\end{equation}

\qquad Next, we need to proceed to the closure of the equations, that is to
express both $\nu_{\mathbf{qQ}}(t)$ and the $2^{nd}$ order fluxes in terms of
the four variables $n$, $h$, $\mathbf{I}_{n}$ and $\mathbf{I}_{h}$.

\qquad Regarding $\nu_{\mathbf{qQ}}(t)$, we apply Heims-Jaynes \cite{heims}
perturbative expansion for averages (around the homogeneous state) in a linear
approximation, to obtain that, for $\mathbf{Q}\neq0$ (see \textbf{Appendix C})%
\begin{equation}
\nu_{\mathbf{qQ}}(t)=b_{1}\left(  \mathbf{q},t\right)  n(\mathbf{Q}%
,t)+b_{2}\left(  \mathbf{q},t\right)  h(\mathbf{Q},t)+\mathbf{b}_{3}\left(
\mathbf{q},t\right)  \cdot\mathbf{I}_{n}(\mathbf{Q},t)+\mathbf{b}_{4}\left(
\mathbf{q},t\right)  \cdot\mathbf{I}_{h}(\mathbf{Q},t). \label{eq43}%
\end{equation}

\qquad Furthermore, developing $\Pi_{\mathbf{q\pm Q/2}}$ around $\mathbf{Q}=0$
and retaining only the lowest order non-vanishing term, that is, taking the
long wavelength limit as indicated in \textbf{Appendix B},%
\begin{equation}
\left(  i/2\right)  \left(  \Pi_{\mathbf{q+Q/2}}-\Pi_{\mathbf{q-Q/2}}\right)
\approx i\mathbf{Q\cdot}\nabla_{\mathbf{q}}\Pi_{\mathbf{q}}, \label{eq44}%
\end{equation}%
\begin{equation}
(1/2)\left(  \Gamma_{\mathbf{q+Q/2}}+\Gamma_{\mathbf{q-Q/2}}\right)
\approx\Gamma_{\mathbf{q}}. \label{eq45}%
\end{equation}

\qquad After introducing Eqs.(\ref{eq43}), (\ref{eq44}) and Eq.(\ref{eq45}) in
Eqs.(\ref{eq39}) to (\ref{eq42}) and, next, going over direct $\mathbf{r}%
$-space we do have that%

\[
\partial n(\mathbf{r},t)/\partial t+\nabla\mathbf{\cdot I}_{n}\mathbf{(r,}%
t\mathbf{)=}\nabla\mathbf{\mathbf{\cdot}}a_{13}^{\left[  2\right]  }\left(
t\right)  \cdot\mathbf{\mathbf{I}_{n}\mathbf{(r,}t\mathbf{)}+}\nabla
\mathbf{\cdot}a_{14}^{\left[  2\right]  }\left(  t\right)  \cdot
\mathbf{\mathbf{I}_{h}\mathbf{(r,}t\mathbf{)}}%
\]%
\begin{equation}
+b_{11}\left(  t\right)  n(\mathbf{r},t)+b_{12}\left(  t\right)
h(\mathbf{r},t)+\mathcal{I}_{n}^{\left[  0\right]  \mathbf{ext.}}\left(
\mathbf{r},t\right)  , \label{eq50}%
\end{equation}%
\[
\partial h(\mathbf{r},t)/\partial t+\nabla\mathbf{\cdot I}_{h}\mathbf{(r,}%
t\mathbf{)=}\nabla\mathbf{\mathbf{\cdot}}a_{24}^{\left[  2\right]  }\left(
t\right)  \cdot\mathbf{\mathbf{I}_{h}\mathbf{(r,}t\mathbf{)}+}\nabla
\mathbf{\cdot}a_{23}^{\left[  2\right]  }\left(  t\right)  \cdot
\mathbf{\mathbf{I}_{n}\mathbf{(r,}t\mathbf{)}}%
\]%
\begin{equation}
+b_{22}\left(  t\right)  h(\mathbf{r},t)+b_{21}\left(  t\right)
n(\mathbf{r},t)+\mathcal{I}_{h}^{\left[  0\right]  \mathbf{ext.}}\left(
\mathbf{r},t\right)  , \label{eq51}%
\end{equation}%
\[
\partial\mathbf{I}_{n}(\mathbf{r},t\mathbf{)/\partial}t+\nabla\mathbf{\cdot
}I_{n}^{\left[  2\right]  }(\mathbf{r},t\mathbf{)=}\nabla\mathbf{\mathbf{\cdot
}}a_{31}^{\left[  2\right]  }\left(  t\right)  n\mathbf{\mathbf{(r,}%
t\mathbf{)}+}\nabla\mathbf{\cdot}a_{32}^{\left[  2\right]  }\left(  t\right)
h\mathbf{\mathbf{(r,}t\mathbf{)}}%
\]%
\begin{equation}
\mathbf{+}b_{33}^{\left[  2\right]  }\left(  t\right)  \cdot\mathbf{\mathbf{I}%
_{n}\mathbf{(r,}t\mathbf{)}+}b_{34}^{\left[  2\right]  }\left(  t\right)
\cdot\mathbf{\mathbf{I}_{h}\mathbf{(r,}t\mathbf{)+}}\mathcal{I}_{n}^{\left[
1\right]  \mathbf{ext.}}\left(  \mathbf{r},t\right)  , \label{eq52}%
\end{equation}%
\[
\partial\mathbf{I}_{h}(\mathbf{r},t\mathbf{)/\partial}t+\nabla\mathbf{\cdot
}I_{h}^{\left[  2\right]  }(\mathbf{r},t\mathbf{)=}\nabla\mathbf{\cdot}%
a_{42}^{\left[  2\right]  }\left(  t\right)  h\mathbf{\mathbf{(r,}%
t\mathbf{)+}}\nabla\mathbf{\mathbf{\cdot}}a_{41}^{\left[  2\right]  }\left(
t\right)  n\mathbf{\mathbf{(r,}t\mathbf{)}}%
\]%
\begin{equation}
\mathbf{+}b_{44}^{\left[  2\right]  }\left(  t\right)  \cdot\mathbf{\mathbf{I}%
_{h}\mathbf{(r,}t\mathbf{)}+}b_{43}^{\left[  2\right]  }\left(  t\right)
\cdot\mathbf{\mathbf{I}_{n}\mathbf{(r,}t\mathbf{)+}}\mathcal{I}_{h}^{\left[
1\right]  \mathbf{ext.}}\left(  \mathbf{r},t\right)  . \label{eq53}%
\end{equation}
The coefficients $a_{ij}^{\left[  2\right]  }\left(  t\right)  $,
$b_{ij}^{\left[  2\right]  }\left(  t\right)  $, which depend on
$\Gamma_{\mathbf{q}}$ and on $\nabla_{\mathbf{q}}\Pi_{\mathbf{q}}$, are given
in \textbf{Appendix D}.

\qquad In order to close the system of equations, one still needs to express
the $2^{nd}$ order fluxes $I_{n}^{\left[  2\right]  }(\mathbf{r},t)$ and
$I_{h}^{\left[  2\right]  }(\mathbf{r},t)$ in terms of the basic variables. We
invoke again Heims-Jaynes \cite{heims} perturbative procedure, in the linear
approximation, to obtain that (see \textbf{Appendix C})
\begin{equation}
\nabla\mathbf{\cdot}I_{n}^{\left[  2\right]  }(\mathbf{r},t\mathbf{)=}%
B_{1}^{\left[  2\right]  }\left(  t\right)  \cdot\nabla n\mathbf{\mathbf{(r,}%
t\mathbf{)}+}B_{2}^{\left[  2\right]  }\left(  t\right)  \cdot\nabla
h\mathbf{\mathbf{(r,}t\mathbf{),}} \label{eq54}%
\end{equation}%
\begin{equation}
\nabla\mathbf{\cdot}I_{h}^{\left[  2\right]  }(\mathbf{r},t\mathbf{)=}%
C_{1}^{\left[  2\right]  }\left(  t\right)  \cdot\nabla n\mathbf{\mathbf{(r,}%
t\mathbf{)}+}C_{2}^{\left[  2\right]  }\left(  t\right)  \cdot\nabla
h\mathbf{\mathbf{(r,}t\mathbf{),}} \label{eq55}%
\end{equation}%
\begin{equation}
B_{j}^{\left[  2\right]  }\left(  t\right)  =%
{\textstyle\sum\nolimits_{\mathbf{q}}}
b_{j}\left(  \mathbf{q},t\right)  \left[  \nabla_{\mathbf{q}}\omega
_{\mathbf{q}}\nabla_{\mathbf{q}}\omega_{\mathbf{q}}\right]  , \label{eq56}%
\end{equation}%
\begin{equation}
C_{j}^{\left[  2\right]  }\left(  t\right)  =%
{\textstyle\sum\nolimits_{\mathbf{q}}}
\hslash\omega_{\mathbf{q}}b_{j}\left(  \mathbf{q},t\right)  \left[
\nabla_{\mathbf{q}}\omega_{\mathbf{q}}\nabla_{\mathbf{q}}\omega_{\mathbf{q}%
}\right]  , \label{eq57}%
\end{equation}
for $j=1$ or $2$, and coefficients $b_{j}\left(  \mathbf{q},t\right)  $ are
given in \textbf{Appendix D}. Using the expressions above for $\nabla
\mathbf{\cdot}I_{n}^{\left[  2\right]  }(\mathbf{r},t\mathbf{)}$ and
$\nabla\mathbf{\cdot}I_{h}^{\left[  2\right]  }(\mathbf{r},t\mathbf{)}$, the
four equations, (\ref{eq50}) to (\ref{eq53}), become%
\[
\partial n(\mathbf{r},t)/\partial t=\left[  \nabla\mathbf{\mathbf{\cdot}%
}a_{13}^{\left[  2\right]  }\left(  t\right)  -\nabla\right]  \mathbf{\cdot
I}_{n}\mathbf{(r,}t\mathbf{)+}\nabla\mathbf{\cdot}a_{14}^{\left[  2\right]
}\left(  t\right)  \cdot\mathbf{\mathbf{I}_{h}\mathbf{(r,}}t\mathbf{\mathbf{)}%
}%
\]%
\begin{equation}
+b_{11}\left(  t\right)  n(\mathbf{r},t)+b_{12}\left(  t\right)
h(\mathbf{r},t)+\mathcal{I}_{n}^{\left[  0\right]  \mathbf{ext.}}\left(
\mathbf{r},t\right)  , \label{eq58}%
\end{equation}%
\[
\partial h(\mathbf{r},t)/\partial t=\left[  \nabla\mathbf{\mathbf{\cdot}%
}a_{24}^{\left[  2\right]  }\left(  t\right)  -\nabla\right]  \mathbf{\cdot
I}_{h}\mathbf{(r,}t\mathbf{)+}\nabla\mathbf{\cdot}a_{23}^{\left[  2\right]
}\left(  t\right)  \cdot\mathbf{\mathbf{I}_{n}\mathbf{(r,}}t\mathbf{\mathbf{)}%
}%
\]%
\begin{equation}
+b_{22}\left(  t\right)  h(\mathbf{r},t)+b_{21}\left(  t\right)
n(\mathbf{r},t)+\mathcal{I}_{h}^{\left[  0\right]  \mathbf{ext.}}\left(
\mathbf{r},t\right)  , \label{eq59}%
\end{equation}%
\[
\partial\mathbf{I}_{n}(\mathbf{r},t\mathbf{)/\partial}t\mathbf{=}\left[
\nabla\mathbf{\mathbf{\cdot}}a_{31}^{\left[  2\right]  }\left(  t\right)
-B_{1}^{\left[  2\right]  }\left(  t\right)  \cdot\nabla\right]
n\mathbf{\mathbf{(r,}}t\mathbf{\mathbf{)}+}\left[  \nabla\mathbf{\cdot}%
a_{32}^{\left[  2\right]  }\left(  t\right)  -B_{2}^{\left[  2\right]
}\left(  t\right)  \cdot\nabla\right]  h\mathbf{\mathbf{(r,}}%
t\mathbf{\mathbf{)}}%
\]%
\begin{equation}
\mathbf{+}b_{33}^{\left[  2\right]  }\left(  t\right)  \cdot\mathbf{\mathbf{I}%
_{n}\mathbf{(r,}}t\mathbf{\mathbf{)}+}b_{34}^{\left[  2\right]  }\left(
t\right)  \cdot\mathbf{\mathbf{I}_{h}\mathbf{(r,}}t\mathbf{\mathbf{)+}%
}\mathcal{I}_{n}^{\left[  1\right]  \mathbf{ext.}}\left(  \mathbf{r},t\right)
, \label{eq60}%
\end{equation}%
\[
\partial\mathbf{I}_{h}(\mathbf{r},t\mathbf{)/\partial}t\mathbf{=}\left[
\nabla\mathbf{\cdot}a_{42}^{\left[  2\right]  }\left(  t\right)
-C_{2}^{\left[  2\right]  }\left(  t\right)  \cdot\nabla\right]
h\mathbf{\mathbf{(r,}}t\mathbf{\mathbf{)+}}\left[  \nabla\mathbf{\mathbf{\cdot
}}a_{41}^{\left[  2\right]  }\left(  t\right)  -C_{1}^{\left[  2\right]
}\left(  t\right)  \cdot\nabla\right]  n\mathbf{\mathbf{(r,}}%
t\mathbf{\mathbf{)}}%
\]%
\begin{equation}
\mathbf{+}b_{44}^{\left[  2\right]  }\left(  t\right)  \cdot\mathbf{\mathbf{I}%
_{h}\mathbf{(r,}}t\mathbf{\mathbf{)}+}b_{43}^{\left[  2\right]  }\left(
t\right)  \cdot\mathbf{\mathbf{I}_{n}\mathbf{(r,}}t\mathbf{\mathbf{)+}%
}\mathcal{I}_{h}^{\left[  1\right]  \mathbf{ext.}}\left(  \mathbf{r},t\right)
. \label{eq61}%
\end{equation}

\qquad The set of equations (\ref{eq58}) to (\ref{eq61}) is a closed system of
four linear first order differential equations for the four
hydro-thermodynamic variables $n(\mathbf{r},t)$, $h(\mathbf{r},t)$,
$\mathbf{I}_{n}(\mathbf{r},t\mathbf{)}$ and $\mathbf{I}_{h}(\mathbf{r}%
,t\mathbf{)}$. Let us now analyze the contents of these equations. In
Eq.(\ref{eq58}) (for the density $n(\mathbf{r},t\mathbf{)}$ of
quasi-particles) and Eq.(\ref{eq59}) (for the density of energy $h(\mathbf{r}%
,t\mathbf{)}$), \textit{the first} term on the RHS is the one of conservation,
that is, the divergence of the corresponding flux; however, they are modified
by the presence of contributions $a_{13}^{\left[  2\right]  }\left(  t\right)
$ and $a_{24}^{\left[  2\right]  }\left(  t\right)  $ arising from the
self-energy correction. The \textit{third term} on the RHS corresponds to a
relaxation-type contribution; the corresponding coefficients $b_{11}\left(
t\right)  $ and $b_{22}\left(  t\right)  $ are minus the inverse of Maxwell
characteristic times which we call $\theta_{n}\left(  t\right)  $ and
$\theta_{h}\left(  t\right)  $. They will be discussed in \textbf{Section IV}.
The second and fourth terms are cross-contributions accounting for
thermo-elastic effects coming both from phonon relaxation effects
(coefficients $b_{12}\left(  t\right)  $ and $b_{21}\left(  t\right)  $) and
phonon energy renormalization (coefficients $a_{14}^{\left[  2\right]
}\left(  t\right)  $ and $a_{23}^{\left[  2\right]  }\left(  t\right)  $). The
last terms $\mathcal{I}_{n}^{\left[  0\right]  \mathbf{ext.}}\left(
\mathbf{r},t\right)  $ and $\mathcal{I}_{h}^{\left[  0\right]  \mathbf{ext.}%
}\left(  \mathbf{r},t\right)  $ account for the rate of change generated by
the externally applied driving agent.

\qquad In Eq.(\ref{eq60}) (for the flux $\mathbf{I}_{n}(\mathbf{r}%
,t\mathbf{)}$ of quasi-particles) and Eq.(\ref{eq61}) (for the flux
$\mathbf{I}_{h}(\mathbf{r},t\mathbf{)}$ of heat), the RHS terms with
coefficient $B_{j}^{\left[  2\right]  }\left(  t\right)  $ and $C_{j}^{\left[
2\right]  }\left(  t\right)  $ had their origin in the divergence of the
$2^{nd}$ order fluxes, which we expressed as superposition of the four basic
hydrodynamic variables. The contributions with $a_{31}^{\left[  2\right]
}\left(  t\right)  $, $a_{32}^{\left[  2\right]  }\left(  t\right)  $,
$a_{42}^{\left[  2\right]  }\left(  t\right)  $ and $a_{41}^{\left[  2\right]
}\left(  t\right)  $ are modifications arising out of self-energy corrections.
The terms with $b_{33}^{\left[  2\right]  }\left(  t\right)  $ and
$b_{44}^{\left[  2\right]  }\left(  t\right)  $, similarly to the cases of
$n(\mathbf{r},t)$ and $h(\mathbf{r},t)$, are relaxation-type contributions
corresponding to minus the inverse of (tensorial) characteristic Maxwell
times, which we call $\theta_{\mathbf{In}}^{\left[  2\right]  }\left(
t\right)  $ and $\theta_{\mathbf{Ih}}^{\left[  2\right]  }\left(  t\right)  $,
see \textbf{Section IV}. As in the equations for $n(\mathbf{r},t)$ and
$h(\mathbf{r},t)$, we also find in the equations for the fluxes cross-terms
proportional to off-diagonal $a^{\left[  2\right]  }$%
\'{}%
s (due to self-energy corrections) and $b^{\left[  2\right]  }$%
\'{}%
s (due to relaxation effects), and external driving forces/sources.

\qquad Furthermore, we call the attention to the fact that in Eq.(\ref{eq60})
for the flux $\mathbf{I}_{n}(\mathbf{r},t\mathbf{)}$ of quasi-particles, the
first term on the RHS (roughly proportional to $\nabla n$) plays the role of a
thermodynamic force analogous in classical fluid hydrodynamics to Fick's Law.
In the same way, the first term on the RHS of Eq.(\ref{eq61}) for the heat
flux $\mathbf{I}_{h}(\mathbf{r},t\mathbf{)}$ (the term here roughly
proportional to $\nabla h$) is analogous to Fourier%
\'{}%
s Law.

\qquad As a matter of fact, Eq.(\ref{eq60}) and Eq.(\ref{eq61}), under
steady-state conditions and after neglecting both external forces/sources and
thermo-elastic cross-terms, reduce to%
\begin{equation}
\mathbf{I}_{n}(\mathbf{r},t\mathbf{)=-}D_{n}^{\left[  2\right]  }\left(
t\right)  \cdot\nabla n(\mathbf{r},t\mathbf{),}\text{ where }D_{n}^{\left[
2\right]  }\left(  t\right)  =\left[  B_{1}^{\left[  2\right]  }\left(
t\right)  -a_{31}^{\left[  2\right]  }\left(  t\right)  \right]  \cdot
\theta_{\mathbf{In}}^{\left[  2\right]  }\left(  t\right)  , \label{eq62}%
\end{equation}%
\begin{equation}
\mathbf{I}_{h}(\mathbf{r},t\mathbf{)=-}D_{h}^{\left[  2\right]  }\left(
t\right)  \cdot\nabla h(\mathbf{r},t\mathbf{),}\text{ where }D_{h}^{\left[
2\right]  }\left(  t\right)  =\left[  C_{2}^{\left[  2\right]  }\left(
t\right)  -a_{42}^{\left[  2\right]  }\left(  t\right)  \right]  \cdot
\theta_{\mathbf{Ih}}^{\left[  2\right]  }\left(  t\right)  . \label{eq63}%
\end{equation}
The \textit{D}'s play the role of rank-2 tensor diffusion coefficients.

\qquad In the next section we analyze several other aspects of this mesoscopic
phonon hydro-thermodynamics of order 1.

\section{Characteristic Maxwell Times and Maxwell-Cattaneo-like Hyperbolic
Equations}

\qquad We consider here some additional characteristics which can be derived
from the results of the previous section.

\subsection{Characteristic Maxwell Times}

\qquad In Eqs.(\ref{eq50})-(\ref{eq53}) the four coefficients $b_{11}$,
$b_{33}^{\left[  2\right]  }$, $b_{22}$, $b_{44}^{\left[  2\right]  }$ are
minus the reciprocal of the so-called Maxwell times \cite{maxwell},
\cite{landau}, namely%
\begin{equation}
-b_{11}\left(  t\right)  \equiv\theta_{n}^{-1}\left(  t\right)  =%
{\textstyle\sum\nolimits_{\mathbf{q}}}
\Gamma_{\mathbf{q}}b_{1}\left(  \mathbf{q},t\right)  =%
{\textstyle\sum\nolimits_{\mathbf{q}}}
w_{\mathbf{n}}\left(  \mathbf{q},t\right)  /\tau_{\mathbf{q}}, \label{eq64}%
\end{equation}%
\begin{equation}
-b_{33}^{\left[  2\right]  }\left(  t\right)  \equiv\left[  \theta
_{\mathbf{In}}^{\left[  -1\right]  }\left(  t\right)  \right]  ^{\left[
2\right]  }=%
{\textstyle\sum\nolimits_{\mathbf{q}}}
\Gamma_{\mathbf{q}}\left[  \mathbf{b}_{3}\left(  \mathbf{q},t\right)
\nabla_{\mathbf{q}}\omega_{\mathbf{q}}\right]  =%
{\textstyle\sum\nolimits_{\mathbf{q}}}
w_{\mathbf{In}}^{\left[  2\right]  }\left(  \mathbf{q},t\right)
/\tau_{\mathbf{q}}, \label{eq65}%
\end{equation}%
\begin{equation}
-b_{22}\left(  t\right)  \equiv\theta_{h}^{-1}\left(  t\right)  =%
{\textstyle\sum\nolimits_{\mathbf{q}}}
\Gamma_{\mathbf{q}}b_{2}\left(  \mathbf{q},t\right)  \hslash\omega
_{\mathbf{q}}=%
{\textstyle\sum\nolimits_{\mathbf{q}}}
w_{\mathbf{h}}\left(  \mathbf{q},t\right)  /\tau_{\mathbf{q}}, \label{eq66}%
\end{equation}%
\begin{equation}
-b_{44}^{\left[  2\right]  }\left(  t\right)  \equiv\left[  \theta
_{\mathbf{Ih}}^{\left[  -1\right]  }\left(  t\right)  \right]  ^{\left[
2\right]  }=%
{\textstyle\sum\nolimits_{\mathbf{q}}}
\Gamma_{\mathbf{q}}\left[  \mathbf{b}_{4}\left(  \mathbf{q},t\right)
\nabla_{\mathbf{q}}\omega_{\mathbf{q}}\right]  \hslash\omega_{\mathbf{q}%
}\text{ }=%
{\textstyle\sum\nolimits_{\mathbf{q}}}
w_{\mathbf{Ih}}^{\left[  2\right]  }\left(  \mathbf{q},t\right)
/\tau_{\mathbf{q}}, \label{eq67}%
\end{equation}
where $\Gamma_{\mathbf{q}}$ is given in Eq.(\ref{eq26}), $b_{j}\left(
\mathbf{q},t\right)  $ for $j=1,2$, $\mathbf{b}_{j}\left(  \mathbf{q}%
,t\right)  $ for $j=3,4$ and $b_{ij}\left(  t\right)  $ are given in
\textbf{Appendix D}. $1/\tau_{\mathbf{q}}=\Gamma_{\mathbf{q}}$ is the
relaxation rate towards the equilibrium phonon population $\nu_{\mathbf{q}%
}^{0}$ in mode $\mathbf{q}$, which depends on $\mathbf{q}$ and $\omega
_{\mathbf{q}}$.

\qquad Eq.(\ref{eq64}) to (\ref{eq67}) tell us that the Maxwell characteristic
times are given by a Mathiessen-like rule involving all the relaxation times
in each mode, $\tau_{\mathbf{q}}$, weighted by different kernels which are
normalized, i.e.,%
\begin{equation}%
{\textstyle\sum\nolimits_{\mathbf{q}}}
w_{\mathbf{n}}\left(  \mathbf{q},t\right)  =%
{\textstyle\sum\nolimits_{\mathbf{q}}}
w_{\mathbf{h}}\left(  \mathbf{q},t\right)  =1, \label{eq68}%
\end{equation}%
\begin{equation}%
{\textstyle\sum\nolimits_{\mathbf{q}}}
w_{\mathbf{In}}^{\left[  2\right]  }\left(  \mathbf{q},t\right)  =%
{\textstyle\sum\nolimits_{\mathbf{q}}}
w_{\mathbf{Ih}}^{\left[  2\right]  }\left(  \mathbf{q},t\right)  =1^{\left[
2\right]  }, \label{eq69}%
\end{equation}
where $1^{\left[  2\right]  }$ is the unit diagonal tensor.

\qquad As a consequence, if we assume all $\tau_{\mathbf{q}}$ to be
independent of $\mathbf{q}$ (not a possible physical situation, see below)
then the Maxwell characteristic times are all equal. On the other hand, if we
apply the mean-value-theorem of calculus, taking outside the $%
{\textstyle\sum\nolimits_{\mathbf{q}}}
$ a suitable mean-value of $\Gamma$ in each case, say $\Gamma^{\left(
\mathbf{n}\right)  }$, $\Gamma^{\left(  \mathbf{h}\right)  }$, $\Gamma
^{\left(  \mathbf{In}\right)  }$ and $\Gamma^{\left(  \mathbf{Ih}\right)  }$,
we find $\theta_{\mathbf{n}}=\tau_{\mathbf{n}}=1/\Gamma^{\left(
\mathbf{n}\right)  }$, $\theta_{\mathbf{h}}=\tau_{\mathbf{h}}=1/\Gamma
^{\left(  \mathbf{h}\right)  }$, $\theta_{\mathbf{In}}^{\left[  2\right]
}=1^{\left[  2\right]  }\tau_{\mathbf{In}}=1^{\left[  2\right]  }%
/\Gamma^{\left(  \mathbf{In}\right)  }$, $\theta_{\mathbf{Ih}}^{\left[
2\right]  }=1^{\left[  2\right]  }\tau_{\mathbf{Ih}}=1^{\left[  2\right]
}/\Gamma^{\left(  \mathbf{Ih}\right)  }$.

\qquad The quantity $\Gamma_{\mathbf{q}}$, given in Eq.(\ref{eq26}), vanishes
unless $\omega_{\mathbf{q}}=$ $\Omega_{\mathbf{k+q}}+\Omega_{\mathbf{k}}$ due
to the $\delta$-function; then it can be rewritten identically as%
\[
\Gamma_{\mathbf{q}}=%
{\textstyle\sum\nolimits_{\mathbf{k}}}
\left\vert M_{\mathbf{kq}}\right\vert ^{2}\delta\left(  \Omega_{\mathbf{k+q}%
}+\Omega_{\mathbf{k}}-\omega_{\mathbf{q}}\right)
\]%
\begin{equation}
\left\{  \left(  e^{\beta\hslash\omega\left(  \mathbf{q}\right)  }-1\right)
/\left\{  e^{\beta\hslash\omega\left(  \mathbf{q}\right)  }\left[
1-e^{-\beta\hslash\omega\left(  \mathbf{q}\right)  }\right]  +\left[
1-e^{\beta\hslash\omega\left(  \mathbf{q}\right)  }\right]  \right\}
\right\}  \label{eq70}%
\end{equation}

\qquad The matrix element $\left\vert M_{\mathbf{kq}}\right\vert ^{2}$ behaves
as $\sim\mathbf{|k+q|}kq$. The frequency $\omega\left(  \mathbf{q}\right)  $
of LA phonons vanishes at the center of the Brillouin zone and is maximum at
the boundary of the zone. Then inspection of Eq.(\ref{eq70}) allows us to
estimate that $\Gamma_{\mathbf{q}}$ is an increasing function of $\left\vert
\mathbf{q}\right\vert $ over the Brillouin zone, therefore, $\tau_{\mathbf{q}%
}=1/\Gamma_{\mathbf{q}}$ is a decreasing one with $\mathbf{q}$ increasing.

\qquad The weighting functions in Eq.(\ref{eq64}) to (\ref{eq67}), and the
time-dependent coefficients $A_{ij}$ and $\Delta_{ij}^{-1}$ below, are given
in \textbf{Appendix D},%
\begin{equation}
w_{\mathbf{n}}\left(  \mathbf{q},t\right)  =\Delta_{12}^{-1}\left(  t\right)
\text{ }\left[  A_{22}\left(  t\right)  -\hslash\omega_{\mathbf{q}}%
A_{12}\left(  t\right)  \right]  \nu_{\mathbf{q}}\left(  t\right)  \left[
1+\nu_{\mathbf{q}}\left(  t\right)  \right]  , \label{eq71}%
\end{equation}%
\begin{equation}
w_{\mathbf{In}}^{\left[  2\right]  }\left(  \mathbf{q},t\right)  =\Delta
_{34}^{-1}\left(  t\right)  \text{ }\left[  \nabla_{\mathbf{q}}\omega
_{\mathbf{q}}\left(  A_{44}^{\left[  2\right]  }\left(  t\right)
-\hslash\omega_{\mathbf{q}}A_{43}^{\left[  2\right]  }\left(  t\right)
\right)  \right]  \cdot\nu_{\mathbf{q}}\left(  t\right)  \left[
1+\nu_{\mathbf{q}}\left(  t\right)  \right]  \nabla_{\mathbf{q}}%
\omega_{\mathbf{q}}, \label{eq72}%
\end{equation}%
\begin{equation}
w_{\mathbf{h}}\left(  \mathbf{q},t\right)  =\Delta_{12}^{-1}\left(  t\right)
\text{ }\left[  A_{11}\left(  t\right)  \hslash\omega_{\mathbf{q}}%
-A_{12}\left(  t\right)  \right]  \nu_{\mathbf{q}}\left(  t\right)  \left[
1+\nu_{\mathbf{q}}\left(  t\right)  \right]  \hslash\omega_{\mathbf{q}},
\label{eq73}%
\end{equation}%
\begin{equation}
w_{\mathbf{Ih}}^{\left[  2\right]  }\left(  \mathbf{q},t\right)  =\Delta
_{34}^{-1}\left(  t\right)  \text{ }\left[  \nabla_{\mathbf{q}}\omega
_{\mathbf{q}}\left(  A_{33}^{\left[  2\right]  }\left(  t\right)
\hslash\omega_{\mathbf{q}}-A_{43}^{\left[  2\right]  }\left(  t\right)
\right)  \right]  \cdot\nu_{\mathbf{q}}\left(  t\right)  \left[
1+\nu_{\mathbf{q}}\left(  t\right)  \right]  \hslash\omega_{\mathbf{q}}%
\nabla_{\mathbf{q}}\omega_{\mathbf{q}}. \label{eq74}%
\end{equation}

In a strictly Debye model it follows that%

\begin{equation}
\theta_{\mathbf{n}}\left(  t\right)  =\theta_{\mathbf{In}}\left(  t\right)  =%
{\textstyle\sum\nolimits_{\mathbf{q}}}
\Gamma_{\mathbf{q}}\nu_{\mathbf{q}}\left(  t\right)  \left[  1+\nu
_{\mathbf{q}}\left(  t\right)  \right]  \left[  \left(  \mathcal{D}_{4}%
-\beta\hslash sq\mathcal{D}_{3}\right)  /\left(  \mathcal{D}_{2}%
\mathcal{D}_{4}-\mathcal{D}_{3}^{2}\right)  \right]  \mathcal{C}^{-1},
\label{eq86}%
\end{equation}%
\begin{equation}
\theta_{\mathbf{h}}\left(  t\right)  =\theta_{\mathbf{Ih}}\left(  t\right)  =%
{\textstyle\sum\nolimits_{\mathbf{q}}}
\Gamma_{\mathbf{q}}\nu_{\mathbf{q}}\left(  t\right)  \left[  1+\nu
_{\mathbf{q}}\left(  t\right)  \right]  \left[  \left(  -\mathcal{D}_{3}%
+\beta\hslash sq\mathcal{D}_{2}\right)  /\left(  \mathcal{D}_{2}%
\mathcal{D}_{4}-\mathcal{D}_{3}^{2}\right)  \right]  \mathcal{C}^{-1}%
\beta\hslash sq, \label{eq87}%
\end{equation}
where $x_{D}=\beta\hslash sq_{D}$, $q_{D}$ is the Debye wavenumber and $s$ is
the sound velocity,
\begin{equation}
\mathcal{D}_{n}=\int_{0}^{x_{D}}dx\text{ }x^{n}\text{ }e^{x}/\left(
e^{x}-1\right)  ^{2} \label{eq88}%
\end{equation}
and%
\begin{equation}
\mathcal{C=}\left(  V/2\pi\right)  \left(  q_{D}/x_{D}\right)  . \label{eq89}%
\end{equation}
It must be stressed that the equalities $\theta_{\mathbf{n}}=\theta
_{\mathbf{In}}$ and $\theta_{\mathbf{h}}=\theta_{\mathbf{Ih}}$ are a
consequence of using a strict Debye model (phonon group velocity independent
of $\mathbf{q}$). This can be a satisfactory approximation only under certain
well-defined restrictions on the macroscopic state of the system.

\qquad Summarizing, the characteristic Maxwell times $\theta$ associated to
the set of fluxes of all orders are composed by the weighted contributions of
the relaxation times of the populations in each mode $\mathbf{q}$, as
described by Eq.(\ref{eq64}) to (\ref{eq67}), which are consistent with
Mathiessen's rule. The $\theta$ are all equal within a strict Debye model, but
we stressed that this approximation is in general too restrictive.

\subsection{Maxwell-Cattaneo-like Hyperbolic Equations}

\qquad Let us consider the four equations, Eq.(\ref{eq58}) to (\ref{eq61}),
where we recall that $b_{11}$, $b_{22}$, $b_{33}$, $b_{44}$ are minus the
reciprocal of Maxwell times [cf. Eqs.(\ref{eq64}) to (\ref{eq67})], and
coefficients $B_{j}^{\left[  2\right]  }$ and $C_{j}^{\left[  2\right]  }$
($j=1,2$) are given in \textbf{Appendix} \textbf{D }( in a Debye model
$\omega\left(  \mathbf{q}\right)  =\omega\left(  q\right)  =sq$,
$B_{2}^{\left[  2\right]  }=C_{1}^{\left[  2\right]  }=0$ and $B_{1}^{\left[
2\right]  }=C_{2}^{\left[  2\right]  }=1^{\left[  2\right]  }s^{2}/3$ ). After
deriving in time Eqs.(\ref{eq58}) and (\ref{eq59}), next introducing
Eqs.(\ref{eq60}) and (\ref{eq61}) and assuming that the kinetic coefficients
are weakly dependent on time, there follow the two coupled $2^{nd}$ order
differential hyperbolic Maxwell-Cattaneo-like equations:%

\[
\partial^{2}n(\mathbf{r},t\mathbf{)/\partial}t^{2}+\left(  \theta
_{\mathbf{In}}^{-1}+\theta_{\mathbf{n}}^{-1}\right)  \partial n(\mathbf{r}%
,t\mathbf{)/\partial}t+\left(  -\nabla\mathbf{\mathbf{\cdot}}B_{1}^{\left[
2\right]  }\left(  t\right)  \cdot\nabla+\theta_{\mathbf{In}}^{-1}%
\theta_{\mathbf{n}}^{-1}+b_{34}b_{21}\right)  n(\mathbf{r},t\mathbf{)}%
\]%
\[
+\left(  b_{34}+b_{21}\right)  \partial h(\mathbf{r},t\mathbf{)/\partial
}t+\left(  -\nabla\mathbf{\mathbf{\cdot}}B_{2}^{\left[  2\right]  }\left(
t\right)  \cdot\nabla+\theta_{\mathbf{In}}^{-1}b_{12}+\theta_{\mathbf{h}}%
^{-1}b_{34}\right)  h(\mathbf{r},t\mathbf{)}%
\]%
\begin{equation}
\mathbf{=}S_{n}\left(  \mathbf{r},t\right)  , \label{eq105}%
\end{equation}%
\[
\partial^{2}h(\mathbf{r},t\mathbf{)/\partial}t^{2}+\left(  \theta
_{\mathbf{Ih}}^{-1}+\theta_{\mathbf{h}}^{-1}\right)  \partial h(\mathbf{r}%
,t\mathbf{)/\partial}t+\left(  -\nabla\mathbf{\mathbf{\cdot}}C_{2}^{\left[
2\right]  }\left(  t\right)  \cdot\nabla+\theta_{\mathbf{Ih}}^{-1}%
\theta_{\mathbf{h}}^{-1}+b_{43}b_{12}\right)  h(\mathbf{r},t\mathbf{)}%
\]%
\[
+\left(  b_{43}+b_{21}\right)  \partial n(\mathbf{r},t\mathbf{)/\partial
}t+\left(  -\nabla\mathbf{\mathbf{\cdot}}C_{1}^{\left[  2\right]  }\left(
t\right)  \cdot\nabla+\theta_{\mathbf{Ih}}^{-1}b_{21}+\theta_{\mathbf{n}}%
^{-1}b_{43}\right)  n(\mathbf{r},t\mathbf{)}%
\]%
\begin{equation}
\mathbf{=}S_{h}\left(  \mathbf{r},t\right)  . \label{eq106}%
\end{equation}
In these equations, the contributions associated with self-energy corrections
were neglected and $S_{n}$ and $S_{h}$ account for sources and/or external forces.

\section{Decoupled Motions of Quasi-particles and Heat}

\qquad Consider again the set of Eq.(\ref{eq58}) to (\ref{eq61}), in real
$\mathbf{r}$ space, for the four thermodynamic variables $n$, $h$,
$\mathbf{I}_{n}$ and $\mathbf{I}_{h}$. From now on we neglect the coefficients
$a_{ij}$, i.e., we disregard contributions arising out of self-energy
corrections, see \textbf{Appendix D}. We also neglect coefficients $b_{ij}$
with $i\neq j$. Such $b_{ij}$ describe thermo-elastic effects, which will be
discussed in Section VI.

\qquad The set of equations Eq.(\ref{eq58}) to (\ref{eq61}) becomes simplified
and quantities related to the family $n$ become decoupled from the quantities
related to the family $h$ and vice-versa, giving the two pairs of equations:%

\begin{equation}
\partial n(\mathbf{r},t)/\partial t\mathbf{+}\nabla\mathbf{\cdot I}%
_{n}\mathbf{(r,}t\mathbf{)=-}\theta_{n}^{-1}\left(  t\right)  n(\mathbf{r}%
,t)+\mathcal{I}_{n}^{\left[  0\right]  \mathbf{ext.}}\left(  \mathbf{r}%
,t\right)  , \label{eq120}%
\end{equation}%
\begin{equation}
\partial\mathbf{I}_{n}(\mathbf{r},t\mathbf{)/\partial}t\mathbf{+}%
B_{1}^{\left[  2\right]  }\left(  t\right)  \cdot\nabla n\mathbf{\mathbf{(r,}%
}t\mathbf{\mathbf{)=}}-\left[  \theta_{\mathbf{In}}^{-1}\left(  t\right)
\right]  ^{\left[  2\right]  }\cdot\mathbf{\mathbf{I}_{n}\mathbf{(r,}%
}t\mathbf{\mathbf{)+}}\mathcal{I}_{n}^{\left[  1\right]  \mathbf{ext.}}\left(
\mathbf{r},t\right)  , \label{eq121}%
\end{equation}%
\begin{equation}
\partial h(\mathbf{r},t)/\partial t+\nabla\mathbf{\cdot I}_{h}\mathbf{(r,}%
t\mathbf{)=}-\theta_{h}^{-1}\left(  t\right)  h(\mathbf{r},t)+\mathcal{I}%
_{h}^{\left[  0\right]  \mathbf{ext.}}\left(  \mathbf{r},t\right)  ,
\label{eq122}%
\end{equation}%
\begin{equation}
\partial\mathbf{I}_{h}(\mathbf{r},t\mathbf{)/\partial}t\mathbf{+}%
C_{2}^{\left[  2\right]  }\left(  t\right)  \cdot\nabla h\mathbf{\mathbf{(r,}%
}t\mathbf{\mathbf{)=}}-\left[  \theta_{\mathbf{Ih}}^{-1}\left(  t\right)
\right]  ^{\left[  2\right]  }\cdot\mathbf{\mathbf{I}_{h}\mathbf{(r,}%
}t\mathbf{\mathbf{)+}}\mathcal{I}_{h}^{\left[  1\right]  \mathbf{ext.}}\left(
\mathbf{r},t\right)  . \label{eq123}%
\end{equation}
The coefficients present in these equations depend on time but not on position
\textbf{r}. Then, taking the time-derivative of Eq.(\ref{eq120}) and the
divergence of Eq.(\ref{eq121}) one can partially eliminate the flux
$\mathbf{I}_{n}$; the intermediate equation for $n(\mathbf{r},t)$ is%
\[
\partial^{2}n(\mathbf{r},t\mathbf{)/\partial}t^{2}-\nabla\cdot\left[
B_{1}^{\left[  2\right]  }\left(  t\right)  \cdot\nabla n\mathbf{\mathbf{(r,}%
}t\mathbf{)}\right]  =-\partial\left[  \theta_{n}^{-1}\left(  t\right)
n(\mathbf{r},t)\right]  /\partial t+\partial\mathcal{I}_{n}^{\left[  0\right]
\mathbf{ext.}}\left(  \mathbf{r},t\right)  /\partial t
\]%
\begin{equation}
+\nabla\cdot\left[  \theta_{\mathbf{In}}^{\left[  -1\right]  }\left(
t\right)  \right]  ^{\left[  2\right]  }\cdot\mathbf{\mathbf{I}_{n}%
\mathbf{(r,}}t\mathbf{\mathbf{)-}}\nabla\mathbf{\mathbf{\cdot}}\mathcal{I}%
_{n}^{\left[  1\right]  \mathbf{ext.}}\left(  \mathbf{r},t\right)  .
\label{eq124}%
\end{equation}
If, in addition, the tensor $\left[  \theta_{\mathbf{In}}^{\left[  -1\right]
}\left(  t\right)  \right]  ^{\left[  2\right]  }$ is isotropic ($\left[
\theta_{\mathbf{In}}^{\left[  -1\right]  }\left(  t\right)  \right]  ^{\left[
2\right]  }=\theta_{\mathbf{In}}^{\left[  -1\right]  }\left(  t\right)
1^{\left[  2\right]  }$), the flux $\mathbf{\mathbf{I}_{n}}$ can be completely
eliminated with help of Eq.(\ref{eq120}), giving%
\[
\partial^{2}n(\mathbf{r},t\mathbf{)/\partial}t^{2}-\nabla\cdot\left[
B_{1}^{\left[  2\right]  }\left(  t\right)  \cdot\nabla n\mathbf{\mathbf{(r,}%
}t\mathbf{)}\right]  =-\partial\left[  \theta_{n}^{-1}\left(  t\right)
n(\mathbf{r},t)\right]  /\partial t+\partial\mathcal{I}_{n}^{\left[  0\right]
\mathbf{ext.}}\left(  \mathbf{r},t\right)  /\partial t
\]%
\begin{equation}
+\theta_{\mathbf{In}}^{\left[  -1\right]  }\left(  t\right)  \left\{
-\partial n(\mathbf{r},t)/\partial t-\theta_{n}^{-1}\left(  t\right)
n(\mathbf{r},t)+\mathcal{I}_{n}^{\left[  0\right]  \mathbf{ext.}}\left(
\mathbf{r},t\right)  \right\}  -\nabla\mathbf{\mathbf{\cdot}}\mathcal{I}%
_{n}^{\left[  1\right]  \mathbf{ext.}}\left(  \mathbf{r},t\right)  .
\label{eq125}%
\end{equation}
If $B_{1}^{\left[  2\right]  }\left(  t\right)  $ is isotropic ($B_{1}%
^{\left[  2\right]  }\left(  t\right)  =B_{1}1^{\left[  2\right]  }$) , the
equation becomes even simpler%
\[
\partial^{2}n(\mathbf{r},t\mathbf{)/\partial}t^{2}-B_{1}\nabla^{2}%
n\mathbf{\mathbf{(r,}}t\mathbf{)=}-\partial\left[  \theta_{n}^{-1}\left(
t\right)  n(\mathbf{r},t)\right]  /\partial t-\theta_{\mathbf{In}}^{\left[
-1\right]  }\left(  t\right)  \partial n(\mathbf{r},t)/\partial t
\]

\begin{equation}
+\theta_{\mathbf{In}}^{\left[  -1\right]  }\left(  t\right)  \mathcal{I}%
_{n}^{\left[  0\right]  \mathbf{ext.}}\left(  \mathbf{r},t\right)
+\partial\mathcal{I}_{n}^{\left[  0\right]  \mathbf{ext.}}\left(
\mathbf{r},t\right)  /\partial t-\nabla\mathbf{\mathbf{\cdot}}\mathcal{I}%
_{n}^{\left[  1\right]  \mathbf{ext.}}\left(  \mathbf{r},t\right)  .
\label{eq126}%
\end{equation}
\ \ \ \ \ \ \ \ \ \ \ \ \ \ \ \ \ The physical interpretation of the terms in
this hyperbolic $2^{nd}$ order partial differential equation for
$n(\mathbf{r},t)$ is as follows: The LHS has the form of a standard
wave-equation in 3 dimensions; the \textit{square propagation speed }is
$B_{1}$. On the RHS we find two damping terms (the ones with time derivatives
of $n(\mathbf{r},t)$) which depend on the two characteristic Maxwell times
$\theta_{n}$ and $\theta_{\mathbf{In}}$, (see \textbf{Section IV}) and a sum
of three other terms which depend on derivatives of the given sources/external
forces, namely $\theta_{\mathbf{In}}^{\left[  -1\right]  }\left(  t\right)
\mathcal{I}_{n}^{\left[  0\right]  \mathbf{ext.}}\left(  \mathbf{r},t\right)
+\partial\mathcal{I}_{n}^{\left[  0\right]  \mathbf{ext.}}\left(
\mathbf{r},t\right)  /\partial t-\nabla\mathbf{\mathbf{\cdot}}\mathcal{I}%
_{n}^{\left[  1\right]  \mathbf{ext.}}\left(  \mathbf{r},t\right)  $. The sum
of these three terms is the \textit{effective source} in Eq.(\ref{eq126}).

\qquad With exactly the same procedures, we get for $h\mathbf{\mathbf{(r,}%
}t\mathbf{)}$ an equation of the same form, but where all $n$ are replaced by
$h$:%
\[
\partial^{2}h(\mathbf{r},t\mathbf{)/\partial}t^{2}-C_{2}\nabla^{2}%
h\mathbf{\mathbf{(r,}}t\mathbf{)=}-\partial\left[  \theta_{h}^{-1}\left(
t\right)  h(\mathbf{r},t)\right]  /\partial t-\theta_{\mathbf{Ih}}^{\left[
-1\right]  }\left(  t\right)  \partial h(\mathbf{r},t)/\partial t
\]%
\begin{equation}
+\theta_{\mathbf{Ih}}^{\left[  -1\right]  }\left(  t\right)  \mathcal{I}%
_{h}^{\left[  0\right]  \mathbf{ext.}}\left(  \mathbf{r},t\right)
+\partial\mathcal{I}_{h}^{\left[  0\right]  \mathbf{ext.}}\left(
\mathbf{r},t\right)  /\partial t-\nabla\mathbf{\mathbf{\cdot}}\mathcal{I}%
_{h}^{\left[  1\right]  \mathbf{ext.}}\left(  \mathbf{r},t\right)  .
\label{eq127}%
\end{equation}

\qquad It can be noticed that these hyperbolic equations resemble the
so-called telegraphist equation in electrodynamics. Regarding the solution of
Eq.(\ref{eq126}) and (\ref{eq127}) we can state that whenever the effective
source has no spectral components at the frequencies of the hydrodynamic
modes, there is a unique \textquotedblleft particular\textquotedblright%
\ solution obtainable with a Green%
\'{}%
s function. Morse and Feshbach \cite{morse} give such Green function for the
simpler case of a strongly localized effective source. However, if the
effective source has spectral components at the frequencies of the
hydrodynamic modes, \textquotedblleft particular\textquotedblright\ solutions
may still be found but they will in general diverge as $|t|\rightarrow\infty$.

\qquad Although the equation for $n(\mathbf{r},t)$, Eq.(\ref{eq126}), and the
equation for $h(\mathbf{r},t)$, Eq.(\ref{eq127}), have the same form, the
coefficients may be widely different in practice. For instance, in
Eq.(\ref{eq126}) for the phonon density $n(\mathbf{r},t)$, the damping (terms
with $\partial n(\mathbf{r},t)/\partial t$) is related to sound attenuation in
the material medium, which is very small in most rigid material media. In this
case, the damping terms are a small correction on the undamped wave equation
and can frequently be ignored.

\qquad Regarding Eq.(\ref{eq127}) for the density of energy, suppose
conditions of quasi-equilibrium have been attained. Then we can define a local
quasi-temperature $T^{\ast}(\mathbf{r},t)$ such that $h(\mathbf{r}%
,t)=C_{V}T^{\ast}(\mathbf{r},t)$ where $C_{V}$ is a constant specific heat per
unit volume. Then, for steady-state processes in ordinary solid material
media, it is empirically established that $\partial^{2}h(\mathbf{r}%
,t\mathbf{)/\partial}t^{2}$ is negligible in comparison with the damping terms
$\partial h(\mathbf{r},t)/\partial t$. In this case, Eq.(\ref{eq127}) reduces
to Fourier%
\'{}%
s equation for $T^{\ast}(\mathbf{r},t)$ \cite{fourier}.

\qquad Let us consider heat motion in more detail. It is governed by
Eq.(\ref{eq122}) and (\ref{eq123}). Deriving Eq.(\ref{eq123}) in time and
taking the spatial gradient of (\ref{eq122}), we eliminate $h(\mathbf{r},t)$
and get the following hyperbolic equation for the heat flux $\mathbf{I}%
_{h}(\mathbf{r},t\mathbf{)}$:%
\[
\partial^{2}\mathbf{I}_{h}(\mathbf{r},t\mathbf{)/\partial}t^{2}+\left(
\theta_{h}^{-1}+\theta_{\mathbf{Ih}}^{-1}\right)  \partial\mathbf{I}%
_{h}(\mathbf{r},t\mathbf{)}/\partial t+\theta_{h}^{-1}\theta_{\mathbf{Ih}%
}^{-1}\mathbf{I}_{h}(\mathbf{r},t\mathbf{)+}C_{2}\nabla\left(  \nabla
\mathbf{\cdot I}_{h}(\mathbf{r},t\mathbf{)}\right)
\]%
\begin{equation}
\mathbf{=}\text{ }\partial\mathcal{I}_{h}^{\left[  1\right]  \mathbf{ext.}%
}\left(  \mathbf{r},t\right)  /\partial t+C_{2}\nabla\mathcal{I}_{h}^{\left[
0\right]  \mathbf{ext.}}\left(  \mathbf{r},t\right)  . \label{eq128}%
\end{equation}
We recall that $\nabla\left(  \nabla\mathbf{\cdot I}_{h}(\mathbf{r}%
,t\mathbf{)}\right)  =\nabla^{2}\mathbf{I}_{h}(\mathbf{r},t\mathbf{)+}%
\nabla\times\nabla\times\mathbf{I}_{h}(\mathbf{r},t\mathbf{)}$. In that way,
Eq.(\ref{eq128}) is an extended version of the Guyer-Krumhansl equation, which
in the steady-state and assuming $\nabla\times\mathbf{I}_{h}(\mathbf{r}%
,t\mathbf{)}=0$ reads as%
\begin{equation}
\mathbf{I}_{h}(\mathbf{r)+}\ell_{h}^{2}\nabla^{2}\mathbf{I}_{h}(\mathbf{r)=}%
\ell_{h}^{2}\nabla\mathcal{I}_{h}^{\left[  0\right]  \mathbf{ext.}}\left(
\mathbf{r}\right)  , \label{eq129}%
\end{equation}
where,%
\begin{equation}
\ell_{h}^{2}=C_{2}\theta_{h}\theta_{\mathbf{Ih}}, \label{eq130}%
\end{equation}
with $\ell_{h}^{2}$ having dimension of length. Notice that in the Debye
model, $C_{2}$ is one third of the square of the sound velocity.

\section{Thermo-elastic Effect}

\qquad When a pulse of energy, well localized in space and time, is absorbed
by a solid, the solid will heat and thermally expand, creating local
time-dependent stresses and strains. As time elapses, the absorbed energy will
spread away from the \textquotedblleft hot spot\textquotedblright\ and the
solid will eventually reach a state of thermal equilibrium with a uniform
temperature, as well as mechanical equilibrium with vanishing stresses and strains.

\qquad Of immediate practical relevance in this analysis are the scalar field
$T^{\ast}(\mathbf{r},t)$ describing a local quasi-temperature, and the vector
field $\mathbf{u}(\mathbf{r},t)$ describing time -and position- dependent
displacements of the atomic nuclei.

Consider a homogeneous equilibrium solid with uniform density $\xi_{0}$. If
now the solid is locally excited in some way and the nuclei in some
neighborhood undergo time dependent displacements, the density, $\xi\left(
r,t\right)  $, will also become time -and position- dependent and for small
displacements can be written as \cite{landau}%
\begin{equation}
\xi\left(  \mathbf{r},t\right)  =\xi_{0}\left[  1-\nabla\cdot\mathbf{u}%
(\mathbf{r},t)\right]  . \label{eq131}%
\end{equation}

\qquad It is clear that all motions of the nuclei can be expressed in terms of
the normal modes of vibration in the solid, or, equivalently, in terms of the
phonon density, be it crystalline or vitreous. Hence we argue that, upon a
time -and space- localized excitation in a solid, the \textit{change} in
material density is, within a multiplicative constant, the same as the
\textit{change} in $n(\mathbf{r},t)$, the phonon-density discussed in the
previous sections. Therefore, we write%
\begin{equation}
\nabla n(\mathbf{r},t)=-\xi_{0}\nabla\left[  \nabla\cdot\mathbf{u}%
(\mathbf{r},t)\right]  , \label{eq132}%
\end{equation}%
\begin{equation}
\partial n(\mathbf{r},t)/\partial t=-\xi_{0}\nabla\cdot\partial\mathbf{u}%
(\mathbf{r},t)/\partial t, \label{eq133}%
\end{equation}%
\begin{equation}
\mathbf{I}_{n}(\mathbf{r},t\mathbf{)}=\xi_{0}\partial\mathbf{u}(\mathbf{r}%
,t)/\partial t. \label{eq134}%
\end{equation}

\qquad If we neglect the off-diagonal coefficients $a_{ij}$ and $b_{ij}$ in
Eq.(\ref{eq50}) and Eq.(\ref{eq52}) and use the definitions of Maxwell
characteristic times (see Eq.(\ref{eq64}) and Eq.(\ref{eq65})), the former
become%
\begin{equation}
\partial n(\mathbf{r},t)/\partial t=-\nabla\mathbf{\cdot I}_{n}\mathbf{(r,}%
t\mathbf{)-}\theta_{n}^{-1}\left(  t\right)  n(\mathbf{r},t)+\mathcal{I}%
_{n}^{\left[  0\right]  \mathbf{ext.}}\left(  \mathbf{r},t\right)  ,
\label{eq135}%
\end{equation}%
\begin{equation}
\partial\mathbf{I}_{n}(\mathbf{r},t\mathbf{)/\partial}t=\mathbf{-}%
\nabla\mathbf{\cdot}I_{n}^{\left[  2\right]  }(\mathbf{r},t)-\theta
_{\mathbf{In}}^{-1}\left(  t\right)  \mathbf{\mathbf{I}_{n}\mathbf{(r,}%
}t\mathbf{\mathbf{)+}}\mathcal{I}_{n}^{\left[  1\right]  \mathbf{ext.}}\left(
\mathbf{r},t\right)  . \label{eq136}%
\end{equation}

\qquad The five equations (\ref{eq132}) through (\ref{eq136}) can be combined
to produce the hyperbolic equation%
\begin{equation}
\xi_{0}\partial^{2}\mathbf{u}(\mathbf{r},t)/\partial t^{2}+\left(  \xi
_{0}/\theta_{\mathbf{In}}\right)  \partial\mathbf{u}(\mathbf{r},t)/\partial
t=\mathbf{-}\nabla\mathbf{\cdot}I_{n}^{\left[  2\right]  }(\mathbf{r}%
,t)\mathbf{\mathbf{+}}\mathcal{I}_{n}^{\left[  1\right]  \mathbf{ext.}}\left(
\mathbf{r},t\right)  . \label{eq137}%
\end{equation}
To close this equation, we need to express $\nabla\mathbf{\cdot}I_{n}^{\left[
2\right]  }(\mathbf{r},t)$ in terms of a mixed representation, what is
described in \textbf{Appendix C. }We end up with%

\[
\frac{\partial^{2}\mathbf{u}\left(  \mathbf{r},t\right)  }{\partial t^{2}%
}+\theta_{\mathbf{In}}^{-1}\left(  t\right)  \frac{\partial\mathbf{u}\left(
\mathbf{r},t\right)  }{\partial t}+\xi_{0}^{-1}A_{nn}^{\left[  2\right]
}\left(  t\right)  \cdot\nabla\left[  \nabla\cdot\mathbf{u}\left(
\mathbf{r},t\right)  \right]  =
\]%
\begin{equation}
=-\frac{\xi_{0}^{-1}}{k_{B}T_{0}^{2}}A_{nh}^{\left[  2\right]  }\left(
t\right)  \cdot\nabla T^{\ast}\left(  \mathbf{r},t\right)  +\xi_{0}%
^{-1}\mathcal{I}_{n}^{\left[  1\right]  \mathbf{ext.}}\left(  \mathbf{r}%
,t\right)  . \label{eq1487}%
\end{equation}

Moreover, as shown in Appendix C, the evolution equation for the
nonequilibrium temperature (quasitemperature $T^{\ast}\left(  \mathbf{r}%
,t\right)  $) is
\[
\theta^{\ast}\frac{\partial^{2}T^{\ast}\left(  \mathbf{r},t\right)  }{\partial
t^{2}}+\frac{\partial T^{\ast}\left(  \mathbf{r},t\right)  }{\partial t}%
+\frac{\theta^{\ast}}{\theta_{h}\theta_{\mathbf{Ih}}}\frac{h(\mathbf{r}%
,t\mathbf{)}}{C_{V}\left(  \mathbf{r},t\right)  }=
\]%
\begin{equation}
=\left[  B_{n}\frac{a_{h}}{a_{n}}-B_{h}\right]  \frac{\theta^{\ast}}%
{k_{B}T^{\ast2}}\frac{1}{C_{V}\left(  \mathbf{r},t\right)  }\nabla^{2}T^{\ast
}\left(  \mathbf{r},t\right)  +\frac{\mathcal{I}_{h}^{\left[  0\right]
\mathbf{ext.}}\left(  \mathbf{r},t\right)  }{C_{V}\left(  \mathbf{r},t\right)
}. \label{eq1488}%
\end{equation}
Where,
\begin{equation}
\frac{1}{\theta^{\ast}}=\frac{1}{\theta_{h}}+\frac{1}{\theta_{\mathbf{Ih}}},
\label{eq1489}%
\end{equation}%
\begin{equation}
B_{n}\frac{a_{h}}{a_{n}}=\frac{%
{\textstyle\sum\nolimits_{\mathbf{q}}}
\hslash\omega_{\mathbf{q}}\nabla_{\mathbf{q}}\omega_{\mathbf{q}}\cdot
\nabla_{\mathbf{q}}\omega_{\mathbf{q}}\text{ }\delta\overline{\nu}%
_{\mathbf{q}}/\delta F_{\mathbf{q}}%
{\textstyle\sum\nolimits_{\mathbf{q}^{,}}}
\hslash\omega_{\mathbf{q}^{,}}\delta\overline{\nu}_{\mathbf{q}^{,}}/\delta
F_{\mathbf{q}^{,}}}{%
{\textstyle\sum\nolimits_{\mathbf{q}}}
\delta\overline{\nu}_{\mathbf{q}}/\delta F_{\mathbf{q}}}, \label{eq1490}%
\end{equation}%
\begin{equation}
B_{h}=%
{\textstyle\sum\nolimits_{\mathbf{q}}}
\left(  \hslash\omega_{\mathbf{q}}\right)  ^{2}\nabla_{\mathbf{q}}%
\omega_{\mathbf{q}}\cdot\nabla_{\mathbf{q}}\omega_{\mathbf{q}}\text{ }%
\delta\overline{\nu}_{\mathbf{q}}/\delta F_{\mathbf{q}}, \label{eq1491}%
\end{equation}
with $\overline{\nu}_{\mathbf{q}}\left(  t\right)  $ given in Eq.(\ref{eqc11}).

\qquad In the evolution equation for $\mathbf{u}(\mathbf{r},t)$
Eq.(\ref{eq1487}) we find in the RHS a term which is proportional to $\nabla
T^{\ast}\left(  \mathbf{r},t\right)  $ in agreement with the phenomenological
equation of elasticity \cite{landau}. In the application to be considered
here, there is no external source of \textquotedblleft particle
flux\textquotedblright\ and we set $\mathcal{I}_{n}^{\left[  1\right]
\mathbf{ext.}}\left(  \mathbf{r},t\right)  =0$. In addition, the term
$\partial\mathbf{u}(\mathbf{r},t)/\partial t$ describes sound attenuation,
which will be negligible in the illustration, hence we drop it.

\qquad In the evolution equation for $T^{\ast}\left(  \mathbf{r},t\right)  $,
Eq.(\ref{eq1488}), there is an external source feeding energy in the solid
medium, namely, a radiant energy pulse, so we keep the term $\mathcal{I}%
_{h}^{\left[  0\right]  \mathbf{ext.}}\left(  \mathbf{r},t\right)  $ to be
further specified below. Except at very short delays after excitation, the
flow of heat is diffusive, which means that $\partial^{2}h(\mathbf{r}%
,t\mathbf{)/\partial}t^{2}\ll\left(  \theta_{h}^{-1}+\theta_{\mathbf{Ih}}%
^{-1}\right)  \partial h(\mathbf{r},t\mathbf{)}/\partial t$, therefore we drop
the $2^{nd}$ order time-derivative. The effect of the $2^{nd}$ order
time-derivative at short delay times was investigated in \cite{spie2011}. The
term linear in $h(\mathbf{r},t)$ describes time-relaxation effects which we
lump together with the effect of the $1^{st}$ order time-derivative. We also
neglect the term with $n(\mathbf{r},t)$, because $n$ is almost uniform and
heat transport is weakly affected by small fluctuations in material density;
in other words, the measured macroscopic diffusion coefficient $D_{h}$ already
contains in itself the effect of density fluctuations which will be always present.

\qquad Given these considerations, we take as practical equations the
following:%
\begin{equation}
\partial T^{\ast}\left(  \mathbf{r},t\right)  /\partial t-D_{h}\nabla
^{2}T^{\ast}\left(  \mathbf{r},t\right)  =C_{V}^{-1}w(\mathbf{r},t),
\label{eq149}%
\end{equation}%
\begin{equation}
\partial^{2}\mathbf{u}(\mathbf{r},t)/\partial t^{2}-c_{s}^{2}\nabla\left(
\nabla\cdot\mathbf{u}(\mathbf{r},t)\right)  =-\left(  \alpha_{V}K/\xi
_{0}\right)  \nabla T^{\ast}\left(  \mathbf{r},t\right)  , \label{eq150}%
\end{equation}
where $D_{h}$ is the diffusion coefficient, $C_{V}$ is the specific heat per
unit volume, $w(\mathbf{r},t)$ is the power density being transferred from the
radiant pulse into the medium, $c_{s}$ is the speed of sound, $\alpha
_{V}=(1/V)(dV/dT)$ is the volumetric coefficient of thermal expansion, $K$ is
the bulk modulus and $\xi_{0}$ is the material density.

\qquad Now we can address, as illustrative application, the transient thermal
distortion of an optical substrate (mirror, Bragg crystal, diffraction
grating, etc...) illuminated by an intense ultra-short X-ray pulse such as
currently available at several Free-Electron-Laser (FEL) facilities in the
world \cite{fel2011}. This situation is entirely different from the case of
optical components subjected to steady-state heat loads.

\qquad The pulses produced at FEL facilities are strongly localized in $3$-dim
space and in time. When such a pulse of X-rays reaches the interface
vacuum/solid at the surface of an optical element, some fraction of the
radiant energy is reflected and some fraction penetrates the solid to a depth
$\delta_{0}$ and is absorbed, generating heat and thermal distortion which
impacts the optical performance of the device. We want to estimate the
seriousness of the surface distortion in a time scale of pico- to nano-seconds
after incidence of a single X-ray pulse lasting only a few femto-seconds. The
analysis below will show that, although the incident femto-second pulse is
gone long before the optical surface has time to distort, the next pulses in a
pulse-train can be badly affected.

\qquad Consider the interface between vacuum at $z<0$ and an infinite slab of
silicon at $z>0$, extending indefinitely in the $x$ and $y$ directions.
Crystalline silicon is a very popular material for optical substrates, because
it is available in large sizes, it has high thermal conductance and low
coefficient of thermal expansion (hence thermal distortion is minimized),
accepts state-of-the-art polishing (RMS surface roughness of only a few
Angstroms) and is relatively cheap.

\qquad A short Gaussian-shaped pulse of X-rays (duration $t_{FEL}$, radius
$r_{0}$) centered at wavelength $\lambda_{FEL}$ and propagating along the $z$
axis comes from $z=-\infty$ and hits the interface at $t=0$. This scenario is
cylindrically symmetric about the $z$ axis. Tables I and II describe the FEL
pulse and the solid medium.

\textbf{Table 1 - FEL characteristics}%

\begin{tabular}
[c]{|l|l|l|l|}\hline\hline
\textit{Quantity} & \textit{Symbol} & \textit{Value} & \textit{Unit}\\\hline
Photon energy & $h\nu=\hslash\omega$ & $5000$ & $eV$\\\hline
Photon wavelength & $\lambda$ & $22.8$ & $\mathring{A}$\\\hline
Pulse duration & $\Delta t_{FEL}$ & $10$ & $fsec$\\\hline
Pulse total energy & $W_{FEL}$ & $0.8$ & $mJ$\\\hline
Peak FEL power & $P_{FEL}$ & $80$ & $GW$\\\hline
\# photons per FEL pulse & $N_{FEL}$ & $1.0\times10^{12}$ & \\\hline
Optical reflectivity & $R$ & $0.8$ & \\\hline
Absorbed energy & $(1-R)W_{FEL}$ & $0.16$ & $mJ$\\\hline
Pulse Gaussian radius & $r_{0}$ & $2.0\times10^{6}$ & $nm$\\\hline
&  & $2.0$ & $mm$\\\hline
\end{tabular}

\bigskip

\textbf{Table 2 - Thermo-elastic constants for silicon}%

\begin{tabular}
[c]{|c|l|l|l|}\hline\hline
\textit{Quantity} & \textit{Symbol} & \textit{Value} & \textit{Unit}\\\hline
Density & $\xi_{0}$ & $2.329$ & $g/cm^{3}$\\\hline
Specific heat (per unit mass) & $C_{m}$ & $0.702$ & $J/(g$ $^{o}C)$\\\hline
Specific heat (per unit volume) & $C_{V}=\xi_{0}C_{m}$ & $1.635$ & $J/(cm^{3}$
$^{o}C)$\\\hline
Thermal Conductivity & $\kappa$ & $1.68$ & $W/\left(  cm\text{ }^{o}C\right)
$\\\hline
Heat diffusion coefficient & $D_{h}$ & $0.102$ & $nm^{2}/fsec$\\\hline
&  & $1.02$ & $cm^{2}/sec$\\\hline
Coeff. of ther. exp. (linear) & $\alpha_{l}$ & $3\times10^{-6}$ & $1/^{o}%
C$\\\hline
Coeff. of ther. exp. (volumetric) & $\alpha_{\text{v}}=3\alpha_{l}$ &
$9\times10^{-6}$ & $1/^{o}C$\\\hline
Elastic (bulk) modulus & $K$ & $1.06\times10^{12}$ & $g/\left(  cm\text{
}sec^{2}\right)  $\\\hline
Shear modulus & $\zeta$ & $2.19\times10^{11}$ & $g/\left(  cm\text{ }%
sec^{2}\right)  $\\\hline
Poisson's ratio & $\eta$ & $0.45$ & \\\hline
Speed of sound & $c_{s}$ & $7.21\times10^{-3}$ & $nm/fsec$\\\hline
&  & $7.21\times10^{5}$ & $cm/sec$\\\hline
Opt. abs. const. ($\hslash\omega=5000$ $eV$) & $\mu$ & $0.555\times10^{-4}$ &
$1/nm$\\\hline
\end{tabular}

\bigskip

\qquad The power density $w(\mathbf{r},t)$ to be used in Eq.(\ref{eq149}) is%
\begin{equation}
w(\mathbf{r},t)=0, \label{eq151}%
\end{equation}
if $z<0$, for any time $-\infty<t<\infty$,
\begin{equation}
w(\mathbf{r},t)=P\left(  x,y,z,t\right)  \exp\left(  -\mu z\right)  ,
\label{eq152}%
\end{equation}
if $z>0$, for any time $-\infty<t<\infty$,%
\begin{equation}
P\left(  x,y,z,t\right)  =U_{0}\pi^{1/2}\left(  z_{0}/t_{FEL}\right)
\exp\left[  -\left(  x^{2}+y^{2}\right)  /r_{0}^{2}\right]  \delta\left(
z-ct\right)  . \label{eq153}%
\end{equation}

\qquad For simplicity we take the limit of \textquotedblleft
short\textquotedblright\ $z_{0}$ and $t_{FEL}$, with $(z_{0}/t_{FEL})=c$, the
speed of light in vacuum. Integrating $P$ over all time and all space we get
the total energy $W_{FEL}=U_{0}$ $\pi^{3/2}r_{0}^{2}z_{0}$ carried by the
pulse, and find that $U_{0}$ has the meaning of average energy density in the
radiant pulse. $\mu$ is the light absorption coefficient in the solid medium,
which, for a given medium, depends on the wavelength $\lambda_{FEL}$ of the
radiation. Notice that the calculation is linear and all results scale
linearly with the amount of absorbed energy, which in this application is
$0.16$ $mJ$.

\qquad The procedure here will be the following. First, one solves
Eq.(\ref{eq1488}) for the temperature field $T^{\ast}(\mathbf{r},t)$. Next,
one uses $\nabla T^{\ast}\left(  \mathbf{r},t\right)  $ as source in
Eq.(\ref{eq1487}) for $\mathbf{u}(\mathbf{r},t)$.

\qquad Green's functions, analytical expressions for the solutions $T^{\ast
}(\mathbf{r},t)$ and $\mathbf{u}(\mathbf{r},t)$, and a discussion of numerical
methods are given in Ref. \cite{rsi2010}. Here, let us recall that
$\mathbf{u}(\mathbf{r},t)=\mathbf{u}^{Part}(\mathbf{r},t)+\mathbf{u}%
^{Free}(\mathbf{r},t)$ where $\mathbf{u}^{Part}$ is a \textit{particular
solution} which depends on the given source, while $\mathbf{u}^{Free}$ is any
arbitrarily chosen \textit{solution of the associated homogeneous equation}
(no sources), chosen to satisfy boundary/initial/asymptotic conditions, as the
case may be. The condition to be met here is that all normal stresses at the
\textquotedblleft free\textquotedblright\ surface $z=0$ of the solid medium be
vanishing: $\sigma_{zi}(z=0)=\sigma_{iz}(z=0)=0$. These conditions lead to
coupled Fredholm integral equations of $1^{st}$ kind, which we have solved
only approximately using truncation, see Ref. \cite{rsi2010}.

\qquad The result of these thermo-elastic calculations is that, on absorption
of $0.16$ $mJ$ of energy from a Gaussian X-ray photon pulse at $\hslash
\omega=5000$ $eV$, with $r_{0}=2.0$ $mm$, lasting $10$ $fsec$, there is an
outwards surface bulge several $nm$ high that comes after the FEL pulse, with
a delay of several hundred nanoseconds.

\qquad Detailed results are shown in Figures 1 to 5.

\qquad Figure 1 shows the temperature $T^{\ast}$ versus depth $z$ inside the
Silicon slab, at selected times after incidence of the X-ray pulse, assuming
$T^{\ast}=0$ $^{o}C$ as initial temperature. The thick vertical line is the
\textquotedblleft causal cut-off\textquotedblright\ at $t=100$ $fsec$; namely,
for $t=100$ $fsec$, the light penetrates only as far as $ct=2.998\times$
$10^{4}$ $nm$, hence the temperature at $\left\vert z\right\vert >ct$ is still
identically zero. The cutoff for the other is off-range in this figure.
However, as $t$ increases, the X-ray pulse penetrates deeper and deeper till
it is depleted by absorption. Then, further heating of Silicon layers far away
from the surface proceeds by diffusion only.%

\begin{figure}[ptb]%
\centering
\includegraphics[
natheight=4.979600in,
natwidth=5.333300in,
height=2.0998in,
width=2.2468in
]%
{MCTJP302.wmf}%
\caption{\textit{Figure-1 \ Temperature }$T^{\ast}$\textit{ versus depth }%
$z$\textit{ inside the solid medium, at }$\rho=0$\textit{, at selected
time-delays }$t$\textit{ after incidence of the pulse, assuming }$T^{\ast
}(r,t<0)=0$\textit{ for all }$r$\textit{.}}%
\end{figure}

\qquad Figure 2 shows the surface displacement of the Silicon slab, at the
center $\rho=0$ of the X-ray light spot, as a function of elapsed time. This
is a contribution of the particular solution. Negative displacement means a
surface bulge. The bulge is maximum at about $400$ $nsec$, and its amplitude
is very large about $6$ $nm$ then goes back to the original position on a time
scale of $msec$.%

\begin{figure}[ptb]%
\centering
\includegraphics[
natheight=4.687300in,
natwidth=5.541700in,
height=2.0219in,
width=2.386in
]%
{MCTJ3T01.wmf}%
\caption{\textit{Figure-2 Surface displacement }$u_{z}^{Part}(r=0,t)$\textit{
versus time-delay }$t$\textit{ after incidence of the pulse.}}%
\end{figure}

\qquad Figure 3 shows the radial dependence of the surface bulge at selected
moments. From the profile at $400$ $nsec$ we find a surface figure error
(maximum slope of the surface) of about $2$ $\mu rad$, which is not small, and
takes microseconds to decay. Notice that the first X-ray pulse hitting
\textquotedblleft cold\textquotedblright\ silicon surface suffers no adverse
effects, because at the time the surface bulges out, the pulse is already long
gone. If, however, the experiment envisages a train of FEL pulses, and the
spacing in the train is a few hundred $nsec$, the pulses following the first
will be defocused by the heat-induced surface bulge. Furthermore, the bulge
can be resonantly enhanced in a disastrous way.%

\begin{figure}[ptb]%
\centering
\includegraphics[
natheight=4.041300in,
natwidth=5.395600in,
height=2.0228in,
width=2.6913in
]%
{MCTK3803.wmf}%
\caption{\textit{Figure-3 Radial profile of the particular solution }%
$u_{z}^{Part}(\rho,z=0,t)$\textit{ at selected time-delays }$t$\textit{ after
incidence of the pulse (logarithmic vertical scale)}}%
\end{figure}

\qquad Figure 4 shows again, for $t=400$ $ns$, the radial profile of the
surface displacement $u_{z}^{Part}$ predicted by the particular solution but
in a linear scale for easy comparison with Figure 5.%

\begin{figure}[ptb]%
\centering
\includegraphics[
natheight=4.197800in,
natwidth=5.521000in,
height=2.0228in,
width=2.6515in
]%
{MCTM7F05.wmf}%
\caption{\textit{Figure-4 Radial profile of the particular solution }%
$u_{z}^{Part}(\rho,z=0,t=400$\textit{ }$ns$\textit{) (linear vertical scale),
for comparison with Figure 5.}}%
\end{figure}

Figure 5 shows, for $t=400$ $ns$, the radial profile of the surface
displacement $u_{z}^{Free}$ predicted by the \ \ \textquotedblleft
free\textquotedblright\ solution. This was obtained by numerical methods,
after truncation of the coupled Fredholm integral equations of $1^{st}$ kind
which follow from the requirement of vanishing normal stresses at the surface
$z=0$. This has opposite sign and is smaller than the particular solution. The
range $0<\rho<7$ $mm$ was cut into several overlapping segments in order to
speed up convergence; different symbols indicate distinct segments. The lack
of perfect overlap at the edges of neighbouring segments gives an estimate of
the errors incurred in the approximate numerical solution of the coupled
integral equations. The net displacement is $u_{z}^{Part}+u_{z}^{Free}=$ $-5$
$nm$ (the minus sign means the displacement is outwards). The slope $du/d\rho$
is known as \textquotedblleft surface figure error\textquotedblright\ and is
an important figure of merit for optical components. Here it is $\sim2$
$10^{-6}$ $rad$, which significantly exceeds the current state-of-the-art in
optical polishing.%

\begin{figure}[ptb]%
\centering
\includegraphics[
natheight=4.177000in,
natwidth=5.458700in,
height=2.0816in,
width=2.7112in
]%
{MCTLWY04.wmf}%
\caption{Figure-5 Radial profile of the free solution $u_{z}^{Free}%
(\rho,z=0,t=400$ $ns$) (linear vertical scale), for comparison with Figure 4.
The range $0<\rho<7$ $mm$ was cut into several overlapping segments indicated
by the differente symbols. The lack of perfect overlap at the edges of
neighbouring segments gives an estimate of the errors incurred in the
approximate numerical solution of the coupled integral equations.}%
\end{figure}

\qquad We can compare the present results to the previously studied case of
vacuum ultra-violet light \cite{rsi2010}, \cite{rsi2011e}, \cite{spie2011}
where the photon energy was $\hslash\omega=100$ $eV$, absorbed energy per unit
area $W_{0}=180$ $\mu J/cm^{2}$, absorption coefficient $0.128$ $\left(
nm\right)  ^{-1}$, peak displacement $u_{Max}=0.02$ $nm$ and \textquotedblleft
specific displacement\textquotedblright\ $u_{Max}/W_{0}=1.3\times10^{-4}$
$nm/(\mu J/cm^{2})$.

\qquad For X-rays, present calculation, we find $u_{Max}/W_{0}=8.4\times
10^{-3}$ $nm/(\mu J/cm^{2})$, which is $65$ times larger, even though the
absorbed energy is comparable. However, if $\mu$ is the wavelength-dependent
absorption coefficient, the penetration depth is of order $1/\mu$. This is the
thickness of the layer that is strongly driven out of equilibrium by
absorption of the radiant energy in the light pulse. It is $7.8$ $nm$ for
vacuum ultra-violet light ($\hslash\omega=100$ $eV$), but $1.8\times10^{4}$
$nm$ for X-rays ($\hslash\omega=5000$ $eV$). We conclude that deep penetration
of light wrecks havoc with the surface of optical substrates.

\section{Concluding Remarks}

\qquad We have presented the description of a complete Mesoscopic
Hydro-Thermodynamics of phonons, which can also be referred to as Higher-Order
Nonlinear Generalized Hydrodynamics of phonons. It significantly extends the
standard hydrodynamics of phonons by introducing a complete description in
terms of the densities of phonon quasi-particles and of the energy,
accompanied with their fluxes of all orders [Cf. Eqs.(\ref{eq10}) and
(\ref{eq16})]. That is, as indicated in the main text, we can talk of the
motion of two families, the one associated to the motion of the quasi-particle
density, together with the evolution equations for its fluxes of all orders,
and the one for the motion of energy density, together with the evolution
equations for its fluxes of all orders [Cf. Eqs. (\ref{eq30}) and
(\ref{eq31})], coupled together by cross-correlations describing
thermo-striction effects.

As already noticed in the introduction MHT allows to cover all kinds of
motion, in that it includes those characterized by short wavelengths and
ultrafast time evolution. The system of coupled equations of motions is
extremely cumbersome, in principle of unmanageable proportions. For handling
it is required, depending on each case being considered, to introduce a
contraction of description implying in retaining only a few number of fluxes,
neglecting those that become negligible in time intervals smaller than the
experimental resolution time. In other words \cite{balian}, the contraction of
description implies in retaining the information considered as relevant for
the problem in hands, disregarding nonrelevant information. In the main text
(Section 2) it has been discussed a criterion for deciding on the order of
contraction. It has been considered in full detail the case of a MHT of order
1, which corresponds to a large class of practical situations. The four
characteristic Maxwell times are evidenced ( they are all important in
determining the order of contraction of description). The coupled hyperbolic
Maxwell-Cattaneo-like equations for the densities of quasiparticles and energy
are derived, and neglecting thermo-striction effects both sets are decoupled
and analyzed, obtaining for the heat transport a Guyer-Krumhansl-like equation.

Finally in Section 6 thermo-striction effects are taken into account to study
the expected thermal distortion in silicon mirrors under incidence of high
intensity X-ray pulses in Free-Electron-Laser facilities, and establishing
limiting conditions.

\begin{acknowledgement}
We acknowledge financial support from S\~{a}o Paulo State Research Foundation
\ (FAPESP). ARV\ and RL are Brazil National Research Council (CNPq) research
fellows. CABS acknowledge a leave of absence granted by the Brazilian
Technological Institute of \ Aeronautics, and is grateful to the Condensed
Matter Physics Department at the University of Campinas for the kind
hospitality there received.
\end{acknowledgement}

\appendix{}

\section{\textbf{The Non-equilibrium Statistical Operator}}

According to NESEF (\cite{akhiezer}, \cite{zubarev}, \cite{ziman} with a short
overview given in \cite{luzzi1}), the nonequilibrium statistical operator in
terms of the basic nonequilibrium variables in sets (\ref{eq5}) and
(\ref{eq7}) is given by%
\begin{equation}
\Re_{\varepsilon}\left(  t\right)  =\varrho_{\varepsilon}\left(  t\right)
\times\varrho_{B},\label{eqA1}%
\end{equation}
where%
\[
\varrho_{\varepsilon}\left(  t\right)  =\exp\left\{  \ln\overline{\varrho
}\left(  t,0\right)  -\int_{-\infty}^{t}dt^{\prime}\text{ }e^{\varepsilon
\left(  t^{\prime}-t\right)  }\right.
\]%
\begin{equation}
\left.  \times\frac{d}{dt^{\prime}}\ln\overline{\varrho}\left(  t^{\prime
},t^{\prime}-t\right)  \right\}  \label{eqA2}%
\end{equation}
with $\overline{\varrho}\left(  t,0\right)  $ being the auxiliary statistical
operator (also called \textquotedblleft instantaneous quasi-equilibrium
operator\textquotedblright) and%
\begin{equation}
\overline{\rho}(t^{\prime},t^{\prime}-t)=exp\{-\phi(t^{\prime})-%
{\textstyle\sum\nolimits_{\mathbf{q}}}
\left[  F_{\mathbf{q}}\left(  t^{\prime}\right)  \widehat{\nu}_{\mathbf{q}%
}(t^{\prime}-t)+%
{\textstyle\sum\nolimits_{\mathbf{Q\neq0}}}
F_{\mathbf{qQ}}\left(  t^{\prime}\right)  \widehat{\nu}_{\mathbf{qQ}%
}(t^{\prime}-t)\right]  \}\label{eqA3}%
\end{equation}
where $t^{\prime}$ is the dependence on time of the non-equilibrium
thermodynamic variables $F$ and the dynamic microvariables, in the Heisenberg
representation, depend on $(t^{\prime}-t)$. Moreover, $\varrho_{B}$ is the
canonical distribution of the bath of acoustic phonons in equilibrium at a
temperature $T_{0}$, and $\phi(t)$ ensuring the normalization condition plays
the role of the logarithm of a nonequilibrium partition function.

\qquad We recall that the second term in the exponent in Eq.(\ref{eqA2})
accounts for historicity and irreversibility in the non-equilibrium state of
the system. The quantity $\varepsilon$ is a positive infinitesimal that goes
to zero after the trace operation in the calculation of averages has been
performed. We also recall that%
\begin{equation}
\varrho_{\varepsilon}\left(  t\right)  =\overline{\varrho}\left(  t,0\right)
+\varrho_{\varepsilon}^{\prime}\left(  t\right)  \label{eqA4}%
\end{equation}
i.e., it has an additive composition property, with a contribution of the
instantaneous quasi-equilibrium statistical operator plus the one of
$\varrho_{\varepsilon}^{\prime}\left(  t\right)  $ which contains the
historicity and produces irreversible evolution.

\qquad Next, consider a change of description consisting into going from the
one in terms of the set (\ref{eq5}) to one in terms of the
(hydrodynamic-in-character) basic microdynamical variables%
\begin{equation}
\left\{  \widehat{n}(\mathbf{r},t),\text{ }\widehat{\mathbf{I}}_{n}%
(\mathbf{r},t\mathbf{),...,}\widehat{I}_{n}^{[\ell]}(\mathbf{r}%
,t\mathbf{),...,}\widehat{h}(\mathbf{r},t),\text{ }\widehat{\mathbf{I}}%
_{h}(\mathbf{r},t\mathbf{),...,}\text{ }\widehat{I}_{h}^{[\ell]}%
(\mathbf{r},t\mathbf{),...}\right\}  , \label{eqA5}%
\end{equation}
whose average values, taken over the non-equilibrium ensemble, of these
operators, are the macrovariables in sets (\ref{eq10}) and (\ref{eq16}), which
we have dubbed as the $n$-family and the $h$-family respectively. In order to
calculate the averages we use Eq.(A.3) once we write%
\[
F_{\mathbf{q}}\left(  \mathbf{r},t\right)  =\varphi_{n}\left(  \mathbf{r}%
,t\right)  +\mathbf{F}_{n}\left(  \mathbf{r},t\right)  \mathbf{\cdot}%
\nabla_{\mathbf{q}}\omega_{\mathbf{q}}+%
{\textstyle\sum\nolimits_{\ell\geqslant2}}
\mathbf{F}_{n}^{\left[  \ell\right]  }\left(  \mathbf{r},t\right)  \odot
\nabla_{\mathbf{q}}^{\left[  \ell\right]  }\omega\mathbf{_{\mathbf{q}}}%
\]%
\begin{equation}
+\varphi_{h}\left(  \mathbf{r},t\right)  \hslash\omega_{\mathbf{q}}%
+\mathbf{F}_{h}\left(  \mathbf{r},t\right)  \mathbf{\cdot}\hslash
\omega_{\mathbf{q}}\nabla_{\mathbf{q}}\omega_{\mathbf{q}}+%
{\textstyle\sum\nolimits_{\ell\geqslant2}}
\mathbf{F}_{h}^{\left[  \ell\right]  }\left(  \mathbf{r},t\right)
\odot\hslash\omega_{\mathbf{q}}\nabla_{\mathbf{q}}^{\left[  \ell\right]
}\omega\mathbf{_{\mathbf{q}}}, \label{eqA6}%
\end{equation}
where $\nabla\mathbf{_{\mathbf{q}}}\omega\mathbf{_{\mathbf{q}}}$ is the group
velocity of phonons with crystal momentum $\mathbf{q}$ and $\nabla
_{\mathbf{q}}^{\left[  \ell\right]  }\omega\mathbf{_{\mathbf{q}}}$ is given in
Eq.(\ref{eq15}). The symbol $\odot$ stands for fully contracted product of tensors.

\qquad\textquotedblleft Contraction of description\textquotedblright\ in a
given order, say $k$, is done by taking as null the quantities $\mathbf{F}%
_{n}^{\left[  \ell\right]  }\left(  \mathbf{r},t\right)  $ and $\mathbf{F}%
_{h}^{\left[  \ell\right]  }\left(  \mathbf{r},t\right)  $ for all $\ell>k$.
In the main text we have introduced a study of MHT of order 1, i. e., keeping
only $\varphi_{n}\left(  \mathbf{r},t\right)  $, $\varphi_{h}\left(
\mathbf{r},t\right)  $, $\mathbf{F}_{n}\left(  \mathbf{r},t\right)  $ and
$\mathbf{F}_{h}\left(  \mathbf{r},t\right)  $, and the closure of the
evolution equations was done using the Heims-Jaynes \cite{heims} method as
described in \textbf{Appendix C}.

\section{\bigskip\textbf{Generalized Peierls-Boltzmann Equation}}

\qquad As indicated in Eq.(\ref{eq23}), the evolution equation for the single
particle distribution $\nu_{\mathbf{qQ}}(t)$, is%
\begin{equation}
\frac{\partial}{\partial t}\nu_{\mathbf{qQ}}(t)=(i\hslash)^{-1}%
Tr\{[\widehat{\nu}_{\mathbf{qQ}},\widehat{H}]\varrho_{\varepsilon}\left(
t\right)  \}. \label{eqB1}%
\end{equation}

\qquad Applying the NESEF-based kinetic theory we find%
\begin{equation}
\partial\nu_{\mathbf{qQ}}(t)/\partial t=\mathcal{J}_{\mathbf{qQ}}^{\left(
0\right)  }(t)+\mathcal{J}_{\mathbf{qQ}}^{\left(  1\right)  }(t)+\mathcal{J}%
_{\mathbf{qQ}}^{\left(  2\right)  }(t), \label{eqB2}%
\end{equation}
where,%
\begin{equation}
\mathcal{J}_{\mathbf{qQ}}^{\left(  0\right)  }(t)=(i\hslash)^{-1}%
Tr\{[\widehat{\nu}_{\mathbf{qQ}},\widehat{H}_{OS}]\varrho\left(  t,0\right)
\}, \label{eqB3}%
\end{equation}%
\begin{equation}
\mathcal{J}_{\mathbf{qQ}}^{\left(  1\right)  }(t)=(i\hslash)^{-1}%
Tr\{[\widehat{\nu}_{\mathbf{qQ}},\left(  \widehat{H}_{SB}+\widehat{H}%
_{SP}\right)  ]\varrho\left(  t,0\right)  \}, \label{eqB4}%
\end{equation}%
\[
\mathcal{J}_{\mathbf{qQ}}^{\left(  2\right)  }(t)=(i\hslash)^{-2}\int%
_{-\infty}^{t}dt^{\prime}e^{\varepsilon\left(  t^{\prime}-t\right)  }%
\]%
\[
Tr\{[\left(  \widehat{H}_{SB}+\widehat{H}_{SP}\right)  ,\left[  \widehat{\nu
}_{\mathbf{qQ}},\left(  \widehat{H}_{SB}\left(  t^{\prime}\right)
+\widehat{H}_{SP}\left(  t^{\prime}\right)  \right)  \right]  ]\varrho\left(
t,0\right)  \}
\]%
\begin{equation}
+(i\hslash)^{-1}Tr\{[\left(  \widehat{H}_{SB}+\widehat{H}_{SP}\right)
,\widehat{\nu}_{\mathbf{qQ}}]\varrho\left(  t,0\right)  \}\left[
\delta\mathcal{J}_{\mathbf{qQ}}^{\left(  1\right)  }(t)/\delta\nu
_{\mathbf{qQ}}\left(  t\right)  \right]  . \label{eqB5}%
\end{equation}
Here, $\delta$ stands for functional derivative. We stress that this
expression corresponds to an approximation where the interactions
$\widehat{H}_{SB}$ and $\widehat{H}_{SP}$ are retained only up to second order
(memory and vertex renormalization effects are neglected). After performing
the calculations it follows that%
\[
\partial\nu_{\mathbf{qQ}}(t)/\partial t=i\left(  \omega_{\mathbf{q+Q/2}%
}-\omega_{\mathbf{q-Q/2}}\right)  \nu_{\mathbf{qQ}}(t)+\text{ }i\left(
\Pi_{\mathbf{q+Q/2}}-\Pi_{\mathbf{q-Q/2}}\right)  \nu_{\mathbf{qQ}}(t)\text{ }%
\]%
\begin{equation}
-\left(  1/2\right)  \left(  \Gamma_{\mathbf{q+Q/2}}+\Gamma_{\mathbf{q-Q/2}%
}\right)  \nu_{\mathbf{qQ}}(t)+\text{ }\mathcal{J}_{\mathbf{qQ}}^{\left(
ext\right)  }(t), \label{eqB6}%
\end{equation}
for $\mathbf{Q\neq0}$, and
\begin{equation}
\partial\nu_{\mathbf{q}}(t)/\partial t=-\Gamma_{\mathbf{q}}\left(
\nu_{\mathbf{q}}(t)-\nu_{\mathbf{q}}^{0}\right)  +\mathcal{J}_{\mathbf{q}%
}^{\left(  ext\right)  }(t), \label{eqB61}%
\end{equation}
for $\mathbf{Q=0}$, where $\nu_{\mathbf{q}}^{0}$ is the distribution in
equilibrium at temperature $T_{0}$. Moreover $\Gamma_{\mathbf{q}}$, the
reciprocal relaxation time of the phonons in mode $\mathbf{q}$, is given in
Eq.(\ref{eq26}) in the main text, and $\Pi_{\mathbf{q}}$, the self-energy
correction of the energy $\omega_{\mathbf{q}}$, is given in Eq.(\ref{eq29}).

\qquad For the sake of simplicity, we introduce an approximation of long
wavelength, i.e., we consider $Q<<Q_{B}$, $Q_{B}$ being the Brillouin radius,
in the form%
\begin{equation}
\left(  \omega_{\mathbf{q+Q/2}}-\omega_{\mathbf{q-Q/2}}\right)  \approx
\mathbf{Q\cdot}\nabla_{\mathbf{q}}\omega_{\mathbf{q}}, \label{eqB7}%
\end{equation}%
\begin{equation}
\left(  \Pi_{\mathbf{q+Q/2}}-\Pi_{\mathbf{q-Q/2}}\right)  \approx
\mathbf{Q\cdot}\nabla_{\mathbf{q}}\Pi_{\mathbf{q}}, \label{eqB8}%
\end{equation}%
\begin{equation}
\left(  \Gamma_{\mathbf{q+Q/2}}+\Gamma_{\mathbf{q-Q/2}}\right)  \approx
2\Gamma_{\mathbf{q}}. \label{eqB9}%
\end{equation}

\qquad Then, when going over to direct space, through the Fourier transform of
variable $\mathbf{Q}$ into $\mathbf{r}$, there follows Eq.(\ref{eq24}) which
has a form resembling the standard Peierls-Boltzmann equation.

\section{Application of Heims-Jaynes Formalism}

\ \ \ \ \ \ Writing for the auxiliary statistical operator (cf. Eq.(\ref{eqA3}%
)
\begin{equation}
\overline{\varrho}\left(  t,0\right)  =\exp\left\{  \widehat{A}+\widehat{B}%
\right\}  /Tr\left\{  \exp\left\{  \widehat{A}+\widehat{B}\right\}  \right\}
, \label{eqc1}%
\end{equation}
where%
\begin{equation}
\widehat{A}=%
{\textstyle\sum\nolimits_{\mathbf{q}}}
F_{\mathbf{q}}\left(  t\right)  \widehat{\nu}_{\mathbf{q}}, \label{eqc2}%
\end{equation}%
\begin{equation}
\widehat{B}=%
{\textstyle\sum\nolimits_{\mathbf{q}}}
{\textstyle\sum\nolimits_{\mathbf{Q}\neq0}}
F_{\mathbf{q}}\left(  \mathbf{Q},t\right)  \widehat{\nu}_{\mathbf{qQ}},
\label{eqc3}%
\end{equation}
that is, a separation in terms of the homogeneous contribution, $\widehat{A}$,
and the departure from it, $\widehat{B}$, and introducing%
\begin{equation}
\overline{\varrho}\left(  t,0\right)  _{0}=\exp\left\{  \widehat{A}\right\}
/Tr\left\{  \exp\left\{  \widehat{A}\right\}  \right\}  ,
\end{equation}
the homogeneous contribution, according to Heims-Jaynes perturbative expansion
for averages \cite{heims} keeping only the first order contribution in the
inhomogeneous part of $\widehat{B}$, and using the contraction in MHT of order
one, i. e. [see Eq.(\ref{eq38})]%
\begin{equation}
F_{\mathbf{q}}\left(  \mathbf{Q},t\right)  =\varphi_{n}\left(  \mathbf{Q}%
,t\right)  +\varphi_{h}\left(  \mathbf{Q},t\right)  \hslash\omega_{\mathbf{q}%
}+\mathbf{F}_{n}\left(  \mathbf{Q},t\right)  \mathbf{\cdot}\nabla_{\mathbf{q}%
}\omega_{\mathbf{q}}+\mathbf{F}_{h}\left(  \mathbf{Q},t\right)  \mathbf{\cdot
}\hslash\omega_{\mathbf{q}}\nabla_{\mathbf{q}}\omega_{\mathbf{q}}%
\end{equation}
it follows that%
\[
\nu_{\mathbf{qQ}}(t)=\left[  \varphi_{n}\left(  \mathbf{Q},t\right)
+\varphi_{h}\left(  \mathbf{Q},t\right)  \hslash\omega_{\mathbf{q}}%
+\mathbf{F}_{n}\left(  \mathbf{Q},t\right)  \mathbf{\cdot}\nabla_{\mathbf{q}%
}\omega_{\mathbf{q}}+\mathbf{F}_{h}\left(  \mathbf{Q},t\right)  \mathbf{\cdot
}\hslash\omega_{\mathbf{q}}\nabla_{\mathbf{q}}\omega_{\mathbf{q}}\right]
\]%
\begin{equation}
\times\nu_{\mathbf{q}}(t)\left[  1+\nu_{\mathbf{q}}(t)\right]  , \label{eqc6}%
\end{equation}
where%
\begin{equation}
\nu_{\mathbf{q}}(t)=\left[  \exp\left\{  F_{\mathbf{q}}\left(  t\right)
\right\}  -1\right]  ^{-1}, \label{eqc7}%
\end{equation}
with
\begin{equation}
F_{\mathbf{q}}\left(  t\right)  =\varphi_{n}\left(  t\right)  +\varphi
_{h}\left(  t\right)  \hslash\omega_{\mathbf{q}}+\mathbf{F}_{n}\left(
t\right)  \mathbf{\cdot}\nabla_{\mathbf{q}}\omega_{\mathbf{q}}+\mathbf{F}%
_{h}\left(  t\right)  \mathbf{\cdot}\hslash\omega_{\mathbf{q}}\nabla
_{\mathbf{q}}\omega_{\mathbf{q}},
\end{equation}
and we notice that%
\begin{equation}
\nu_{\mathbf{q}}(t)\left[  1+\nu_{\mathbf{q}}(t)\right]  =\delta
\nu_{\mathbf{q}}(t)/\delta F_{\mathbf{q}}\left(  t\right)  .
\end{equation}

Moreover, we write%
\begin{equation}
\nu_{\mathbf{q}}(t)\simeq\overline{\nu}_{\mathbf{q}}(t)-\overline{\nu
}_{\mathbf{q}}(t)\left[  1+\overline{\nu}_{\mathbf{q}}(t)\right]  \left[
\mathbf{F}_{n}\left(  t\right)  \mathbf{\cdot}\nabla_{\mathbf{q}}%
\omega_{\mathbf{q}}+\mathbf{F}_{h}\left(  t\right)  \mathbf{\cdot}%
\hslash\omega_{\mathbf{q}}\nabla_{\mathbf{q}}\omega_{\mathbf{q}}\right]  ,
\label{eqc10}%
\end{equation}
where%
\begin{equation}
\overline{\nu}_{\mathbf{q}}(t)=\left[  \exp\left\{  \varphi_{n}\left(
t\right)  +\varphi_{h}\left(  t\right)  \hslash\omega_{\mathbf{q}}\right\}
-1\right]  ^{-1}, \label{eqc11}%
\end{equation}
that is, a first order Taylor expansion in $\mathbf{F}_{n}$ and $\mathbf{F}%
_{h}$ (linear approximation).

Next, resorting to the use of the nonequilibrium equations of state that
relate the four nonequilibrium thermodynamic variables in Eq.(\ref{eqc6}) to
the four basic variables, i. e., in reciprocal space
\begin{equation}
n\left(  \mathbf{Q},t\right)  =%
{\textstyle\sum\nolimits_{\mathbf{q}}}
\nu_{\mathbf{qQ}}(t)=\overline{A}_{11}\left(  t\right)  \varphi_{n}\left(
\mathbf{Q},t\right)  +\overline{A}_{12}\left(  t\right)  \varphi_{h}\left(
\mathbf{Q},t\right)  , \label{eqc12}%
\end{equation}%
\begin{equation}
\mathbf{I}_{n}\left(  \mathbf{Q},t\right)  =\overline{A}_{33}^{\left[
2\right]  }\left(  t\right)  \cdot\mathbf{F}_{n}\left(  \mathbf{Q},t\right)
+\overline{A}_{34}^{\left[  2\right]  }\left(  t\right)  \cdot\mathbf{F}%
_{h}\left(  \mathbf{Q},t\right)  , \label{eqc13}%
\end{equation}%
\begin{equation}
h\left(  \mathbf{Q},t\right)  =\overline{A}_{12}\left(  t\right)  \varphi
_{n}\left(  \mathbf{Q},t\right)  +\overline{A}_{22}\left(  t\right)
\varphi_{h}\left(  \mathbf{Q},t\right)  , \label{eqc14}%
\end{equation}%
\begin{equation}
\mathbf{I}_{h}\left(  \mathbf{Q},t\right)  =\overline{A}_{34}^{\left[
2\right]  }\left(  t\right)  \cdot\mathbf{F}_{n}\left(  \mathbf{Q},t\right)
+\overline{A}_{44}^{\left[  2\right]  }\left(  t\right)  \cdot\mathbf{F}%
_{h}\left(  \mathbf{Q},t\right)  , \label{eqc15}%
\end{equation}
where $\overline{A}_{11}$, $\overline{A}_{12}$, $\overline{A}_{33}^{\left[
2\right]  }$, $\overline{A}_{34}^{\left[  2\right]  }$, $\overline{A}_{12}$,
$\overline{A}_{22}$, $\overline{A}_{34}^{\left[  2\right]  }$ and
$\overline{A}_{44}^{\left[  2\right]  }$ are those of Eqs.(\ref{eqd5}),
(\ref{eqd6}), (\ref{eqd8}), (\ref{eqd9}), (\ref{eqd5}), (\ref{eqd7}),
(\ref{eqd9}) and (\ref{eqd10}) in Appendix D, except for the replacement of
$\nu_{\mathbf{q}}(t)$ of Eq.(\ref{eqc10}) by $\overline{\nu}_{\mathbf{q}}(t)$
of Eq.(\ref{eqc11}).

In Eqs.(\ref{eqc12}) and (\ref{eqc14}) the contributions in $\mathbf{F}_{n}$
and $\mathbf{F}_{h}$ present in Eq.(\ref{eqc10}) are null, whereas in
Eqs.(\ref{eqc13}) and (\ref{eqc15}) are null the contributions in $\varphi
_{n}$ and $\varphi_{h}$. Eqs.(\ref{eqc12}) to (\ref{eqc15}) constitute a set
of linear algebraic equations that can be inverted to obtain the four
nonequilibrium thermodynamic variables $\varphi_{n}$, $\varphi_{h}$,
$\mathbf{F}_{n}$ and $\mathbf{F}_{h}$, in terms of the basic hydrodynamic
quantities, $n$, $h$, $\mathbf{I}_{n}$ and $\mathbf{I}_{h}$.

The second-order fluxes are given by%
\begin{equation}
I_{n}^{\left[  2\right]  }(\mathbf{Q},t\mathbf{)=}\overline{A}_{33}^{\left[
2\right]  }\left(  t\right)  \varphi_{n}\left(  \mathbf{Q},t\right)
+\overline{A}_{34}^{\left[  2\right]  }\left(  t\right)  \varphi_{h}\left(
\mathbf{Q},t\right)  , \label{eqc16}%
\end{equation}

\begin{equation}
I_{h}^{\left[  2\right]  }(\mathbf{Q},t\mathbf{)=}\overline{A}_{34}^{\left[
2\right]  }\left(  t\right)  \varphi_{n}\left(  \mathbf{Q},t\right)
+\overline{A}_{44}^{\left[  2\right]  }\left(  t\right)  \varphi_{h}\left(
\mathbf{Q},t\right)  , \label{eqc17}%
\end{equation}
where $\overline{A}_{33}^{\left[  2\right]  }$, $\overline{A}_{34}^{\left[
2\right]  }$ and $\overline{A}_{44}^{\left[  2\right]  }$ are those of
Eqs.(\ref{eqd8}), (\ref{eqd9}) and (\ref{eqd10}) in Appendix D, except for the
replacement of $\nu_{\mathbf{q}}(t)$ of Eq.(\ref{eqc10}) by $\overline{\nu
}_{\mathbf{q}}(t)$ of Eq.(\ref{eqc11}).

Using the nonequilibrium equations of state, after going over direct space we
arrive at the expressions for the divergence of both second-order fluxes given
in Eqs.(\ref{eq54}) and (\ref{eq55}), and then to the closed system of
Eqs.(\ref{eq58}) to (\ref{eq61}).

On the other hand, introducing the concept of nonequilibrium temperature,
better called quasitemperature $T^{\ast}\left(  \mathbf{r},t\right)  $ in the
form%
\begin{equation}
k_{B}T^{\ast}\left(  \mathbf{r},t\right)  =\varphi_{h}^{-1}\left(
\mathbf{r},t\right)  , \label{eqc18}%
\end{equation}
we can obtain an evolution equation for it starting with the evolution
equation for the energy in the form of the hyperbolic Maxwell-Cattaneo
equation, Eq.(\ref{eq106}), from which together with the nonequilibrium
thermodynamic equation of state, Eq.(\ref{eqc14}), we have that
\[
\left[
{\textstyle\sum\nolimits_{\mathbf{q}}}
\left(  \hslash\omega_{\mathbf{q}}\right)  ^{2}\overline{\nu}_{\mathbf{q}%
}(t)\left[  1+\overline{\nu}_{\mathbf{q}}(t)\right]  \right]  \left[
\frac{\partial^{2}\varphi_{h}\left(  \mathbf{r},t\right)  }{\partial t^{2}%
}+\left(  \theta_{h}^{-1}+\theta_{\mathbf{Ih}}^{-1}\right)  \frac
{\partial\varphi_{h}\left(  \mathbf{r},t\right)  }{\partial t}\right]
\]

\[
+\frac{h(\mathbf{r},t\mathbf{)}}{\theta_{h}\theta_{\mathbf{Ih}}}=\nabla
\cdot\left[  \overline{A}_{43}^{\left[  2\right]  }\left(  t\right)
\frac{\overline{A}_{12}\left(  t\right)  }{\overline{A}_{11}\left(  t\right)
}-\overline{A}_{44}^{\left[  2\right]  }\left(  t\right)  \right]  \cdot
\nabla\varphi_{h}\left(  \mathbf{r},t\right)
\]%
\begin{equation}
+\nabla\cdot\left[  \frac{\overline{A}_{34}^{\left[  2\right]  }\left(
t\right)  }{\overline{A}_{11}\left(  t\right)  }\right]  \cdot\nabla n\left(
\mathbf{r},t\right)  +\theta_{\mathbf{Ih}}^{-1}\mathcal{I}_{h}^{\left[
0\right]  \mathbf{ext.}}\left(  \mathbf{r},t\right)  , \label{eqc19}%
\end{equation}
and, after introducing the heat capacity
\begin{equation}
C_{V}\left(  t\right)  =k_{B}%
{\textstyle\sum\nolimits_{\mathbf{q}}}
\left(  \frac{\hslash\omega_{\mathbf{q}}}{k_{B}T_{0}}\right)  ^{2}%
\overline{\nu}_{\mathbf{q}}(t)\left[  1+\overline{\nu}_{\mathbf{q}}(t)\right]
, \label{eqc20}%
\end{equation}
where $T_{0}$ is the temperature in equilibrium in this linear treatment, we
arrive at Eq.(\ref{eq1488}).

\section{NESEF Kinetic Theory and Expressions of the Coefficients in
Eqs.(\ref{eq43}) and (\ref{eq50}) to (\ref{eq61})}

The several coefficients for $\nu_{\mathbf{qQ}}(t)$ in Eq.(\ref{eq43}) are as
follows:%
\begin{equation}
b_{1}(\mathbf{q},t)=\Delta_{12}^{-1}\left(  t\right)  \text{ }\left[
A_{22}\left(  t\right)  -\hslash\omega_{\mathbf{q}}A_{12}\left(  t\right)
\right]  \nu_{\mathbf{q}}\left(  t\right)  \left[  1+\nu_{\mathbf{q}}\left(
t\right)  \right]  , \label{eqd1}%
\end{equation}%
\begin{equation}
b_{2}(\mathbf{q},t)=\Delta_{12}^{-1}\left(  t\right)  \text{ }\left[
A_{11}\left(  t\right)  \hslash\omega_{\mathbf{q}}-A_{12}\left(  t\right)
\right]  \nu_{\mathbf{q}}\left(  t\right)  \left[  1+\nu_{\mathbf{q}}\left(
t\right)  \right]  , \label{eqd2}%
\end{equation}%
\begin{equation}
\mathbf{b}_{3}(\mathbf{q},t)=\Delta_{34}^{-1}\left(  t\right)  \left[
A_{44}^{\left[  2\right]  }\left(  t\right)  -\hslash\omega_{\mathbf{q}}%
A_{34}^{\left[  2\right]  }\left(  t\right)  \right]  \cdot\nu_{\mathbf{q}%
}\left(  t\right)  \text{ }\left[  1+\nu_{\mathbf{q}}\left(  t\right)
\right]  \nabla\mathbf{_{\mathbf{q}}}\omega\mathbf{_{\mathbf{q}}},
\label{eqd3}%
\end{equation}%
\begin{equation}
\mathbf{b}_{4}(\mathbf{q},t)=\Delta_{34}^{-1}\left(  t\right)  \left[
A_{33}^{\left[  2\right]  }\left(  t\right)  \hslash\omega_{\mathbf{q}}%
-A_{34}^{\left[  2\right]  }\left(  t\right)  \right]  \cdot\nu_{\mathbf{q}%
}\left(  t\right)  \text{ }\left[  1+\nu_{\mathbf{q}}\left(  t\right)
\right]  \nabla\mathbf{_{\mathbf{q}}}\omega\mathbf{_{\mathbf{q}}},
\label{eqd4}%
\end{equation}

where the quantities $A_{ij}$ and $\Delta_{ij}$ are,%
\begin{equation}
A_{11}\left(  t\right)  =%
{\textstyle\sum\nolimits_{\mathbf{q}}}
\nu_{\mathbf{q}}\left(  t\right)  \left[  1+\nu_{\mathbf{q}}\left(  t\right)
\right]  , \label{eqd5}%
\end{equation}%
\begin{equation}
A_{12}\left(  t\right)  =A_{21}\left(  t\right)  =%
{\textstyle\sum\nolimits_{\mathbf{q}}}
\nu_{\mathbf{q}}\left(  t\right)  \left[  1+\nu_{\mathbf{q}}\left(  t\right)
\right]  \hslash\omega_{\mathbf{q}}, \label{eqd6}%
\end{equation}%
\begin{equation}
A_{22}\left(  t\right)  =%
{\textstyle\sum\nolimits_{\mathbf{q}}}
\nu_{\mathbf{q}}\left(  t\right)  \left[  1+\nu_{\mathbf{q}}\left(  t\right)
\right]  \left(  \hslash\omega_{\mathbf{q}}\right)  ^{2}, \label{eqd7}%
\end{equation}%
\begin{equation}
A_{33}^{\left[  2\right]  }\left(  t\right)  =%
{\textstyle\sum\nolimits_{\mathbf{q}}}
\nu_{\mathbf{q}}\left(  t\right)  \left[  1+\nu_{\mathbf{q}}\left(  t\right)
\right]  \left[  \nabla\mathbf{_{\mathbf{q}}}\omega\mathbf{_{\mathbf{q}}%
}\nabla\mathbf{_{\mathbf{q}}}\omega\mathbf{_{\mathbf{q}}}\right]  ,
\label{eqd8}%
\end{equation}%
\begin{equation}
A_{34}^{\left[  2\right]  }\left(  t\right)  =A_{43}^{\left[  2\right]
}\left(  t\right)  =%
{\textstyle\sum\nolimits_{\mathbf{q}}}
\nu_{\mathbf{q}}\left(  t\right)  \left[  1+\nu_{\mathbf{q}}\left(  t\right)
\right]  \left[  \nabla\mathbf{_{\mathbf{q}}}\omega\mathbf{_{\mathbf{q}}%
}\nabla\mathbf{_{\mathbf{q}}}\omega\mathbf{_{\mathbf{q}}}\right]
\hslash\omega_{\mathbf{q}}, \label{eqd9}%
\end{equation}%
\begin{equation}
A_{44}^{\left[  2\right]  }\left(  t\right)  =%
{\textstyle\sum\nolimits_{\mathbf{q}}}
\nu_{\mathbf{q}}\left(  t\right)  \left[  1+\nu_{\mathbf{q}}\left(  t\right)
\right]  \left[  \nabla\mathbf{_{\mathbf{q}}}\omega\mathbf{_{\mathbf{q}}%
}\nabla\mathbf{_{\mathbf{q}}}\omega\mathbf{_{\mathbf{q}}}\right]  \left(
\hslash\omega_{\mathbf{q}}\right)  ^{2}, \label{eqd10}%
\end{equation}%
\begin{equation}
\Delta_{12}\left(  t\right)  =A_{11}\left(  t\right)  A_{22}\left(  t\right)
-A_{12}\left(  t\right)  A_{12}\left(  t\right)  \label{eqd11}%
\end{equation}%
\begin{equation}
\Delta_{34}\left(  t\right)  =A_{33}^{\left[  2\right]  }\left(  t\right)
\odot A_{44}^{\left[  2\right]  }\left(  t\right)  -A_{34}^{\left[  2\right]
}\left(  t\right)  \odot A_{34}^{\left[  2\right]  }\left(  t\right)  .
\label{eqd12}%
\end{equation}
In these expressions, $\left[  \nabla\mathbf{_{\mathbf{q}}}\omega
\mathbf{_{\mathbf{q}}}\nabla\mathbf{_{\mathbf{q}}}\omega\mathbf{_{\mathbf{q}}%
}\right]  $ denotes the second order tensor with components $\partial
\omega/\partial q_{i}\partial\omega/\partial q_{j}$, while $F^{\left[
2\right]  }\odot G^{\left[  2\right]  }$ = $%
{\textstyle\sum\nolimits_{ij}}
F_{ij}G_{ji}$, and%
\begin{equation}
\nu_{\mathbf{q}}(t)=\left[  \exp\left\{  F_{\mathbf{q}}\left(  t\right)
\right\}  -1\right]  ^{-1} \label{eqd13}%
\end{equation}
is the population in mode $\mathbf{q}$ (see \textbf{Appendix C}), and finally%

\begin{equation}
a_{13}^{\left[  2\right]  }\left(  t\right)  =%
{\textstyle\sum\nolimits_{\mathbf{q}}}
\left[  \mathbf{b}_{3}(\mathbf{q},t)\nabla_{\mathbf{q}}\Pi_{\mathbf{q}%
}\right]  , \label{eqd14}%
\end{equation}%
\begin{equation}
a_{14}^{\left[  2\right]  }\left(  t\right)  =%
{\textstyle\sum\nolimits_{\mathbf{q}}}
\left[  \mathbf{b}_{4}(\mathbf{q},t)\nabla_{\mathbf{q}}\Pi_{\mathbf{q}%
}\right]  , \label{eqd15}%
\end{equation}%
\begin{equation}
a_{23}^{\left[  2\right]  }\left(  t\right)  =%
{\textstyle\sum\nolimits_{\mathbf{q}}}
\hslash\omega_{\mathbf{q}}\left[  \mathbf{b}_{3}(\mathbf{q},t)\nabla
_{\mathbf{q}}\Pi_{\mathbf{q}}\right]  , \label{eqd16}%
\end{equation}%
\begin{equation}
a_{24}^{\left[  2\right]  }\left(  t\right)  =%
{\textstyle\sum\nolimits_{\mathbf{q}}}
\hslash\omega_{\mathbf{q}}\left[  \mathbf{b}_{4}(\mathbf{q},t)\nabla
_{\mathbf{q}}\Pi_{\mathbf{q}}\right]  , \label{eqd17}%
\end{equation}%
\begin{equation}
b_{11}\left(  t\right)  =-%
{\textstyle\sum\nolimits_{\mathbf{q}}}
b_{1}(\mathbf{q},t)\Gamma_{\mathbf{q}}, \label{eqd18}%
\end{equation}%
\begin{equation}
b_{12}\left(  t\right)  =-%
{\textstyle\sum\nolimits_{\mathbf{q}}}
b_{2}(\mathbf{q},t)\Gamma_{\mathbf{q}}, \label{eqd19}%
\end{equation}%
\begin{equation}
b_{21}\left(  t\right)  =-%
{\textstyle\sum\nolimits_{\mathbf{q}}}
b_{1}(\mathbf{q},t)\Gamma_{\mathbf{q}}\hslash\omega_{\mathbf{q}},
\label{eqd20}%
\end{equation}%
\begin{equation}
b_{22}\left(  t\right)  =-%
{\textstyle\sum\nolimits_{\mathbf{q}}}
b_{2}(\mathbf{q},t)\Gamma_{\mathbf{q}}\hslash\omega_{\mathbf{q}},
\label{eqd21}%
\end{equation}%
\begin{equation}
b_{33}^{\left[  2\right]  }\left(  t\right)  =-%
{\textstyle\sum\nolimits_{\mathbf{q}}}
\left[  \nabla\mathbf{\mathbf{_{\mathbf{q}}}}\omega
\mathbf{\mathbf{_{\mathbf{q}}}b}_{3}(\mathbf{q},t)\right]  \Gamma_{\mathbf{q}%
}, \label{eqd22}%
\end{equation}%
\begin{equation}
b_{34}^{\left[  2\right]  }\left(  t\right)  =-%
{\textstyle\sum\nolimits_{\mathbf{q}}}
\left[  \nabla\mathbf{\mathbf{\mathbf{_{\mathbf{q}}}}}\omega
\mathbf{\mathbf{\mathbf{_{\mathbf{q}}}}b}_{4}(\mathbf{q},t)\right]
\Gamma_{\mathbf{q}}, \label{eqd23}%
\end{equation}%
\begin{equation}
b_{43}^{\left[  2\right]  }\left(  t\right)  =-%
{\textstyle\sum\nolimits_{\mathbf{q}}}
\left[  \nabla\mathbf{\mathbf{\mathbf{_{\mathbf{q}}}}}\omega
\mathbf{\mathbf{\mathbf{_{\mathbf{q}}}}b}_{3}(\mathbf{q},t)\right]
\Gamma_{\mathbf{q}}\hslash\omega_{\mathbf{q}}, \label{eqd24}%
\end{equation}%
\begin{equation}
b_{44}^{\left[  2\right]  }\left(  t\right)  =-%
{\textstyle\sum\nolimits_{\mathbf{q}}}
\left[  \nabla\mathbf{\mathbf{\mathbf{_{\mathbf{q}}}}}\omega
\mathbf{\mathbf{\mathbf{_{\mathbf{q}}}}b}_{4}(\mathbf{q},t)\right]
\Gamma_{\mathbf{q}}\hslash\omega_{\mathbf{q}}, \label{eqd25}%
\end{equation}
and we recall that we are writing $\left[  \mathbf{AB}\right]  $ for the
tensorial product of vectors $\mathbf{A}$ and $\mathbf{B}$, rendering a tensor
of order two.

\end{document}